\documentclass[aps,floatfix,twocolumn,superscriptaddress]{revtex4-1}
\usepackage{xcolor} 
\usepackage{graphicx}
\usepackage[hidelinks]{hyperref}
\usepackage{amssymb}
\usepackage[utf8]{inputenc}
\usepackage{amsmath}

\newcommand{\mat}[1]{\mathrm{#1}}

\renewcommand{\vec}[1]{\boldsymbol{#1}}

\usepackage[hidelinks]{hyperref}
\definecolor{vastkust}{RGB}{0, 48, 80} 
\definecolor{morkkoppar}{RGB}{0, 108, 92}  
\hypersetup{
  colorlinks,
  citecolor=vastkust,
  linkcolor=vastkust,
  urlcolor=vastkust}

\newcommand{\change}[1]{\textcolor{black}{#1}}

\begin{document}
\title{Composite quadrupole order in ferroic and multiferroic materials}
\author{R.~Matthias\ Geilhufe}
\email{matthias.geilhufe@chalmers.se}
\affiliation{Department of Physics, Chalmers University of Technology, 412 96 G\"{o}teborg, Sweden}

\date{\today}
\begin{abstract}
The formalism of composite and intertwined orders has been remarkably successful in discussing the complex phase diagrams of strongly correlated materials and high-$T_c$ superconductors. Here, we propose that composite orders are also realized in ferroelectric and ferromagnetic materials when lattice anisotropy is taken into account. This composite order emerges above the ferroic phase transition, and its type is determined by the easy axis of magnetization or polarization, respectively. In multiferroic materials, where polarization and magnetization are coupled, composites of both orders are possible. This formalism of composite orders naturally accounts for magnetoelectric monopole, toroidal, and quadrupole orders. More broadly, composite orders may explain precursor phenomena in incipient ferroic materials, arising at temperatures above the ferroic phase transition and potentially contributing to the characterization of currently hidden orders.
\end{abstract}
\maketitle
\section{Introduction}
Condensed matter consists of a multitude of ions and electrons, often arranging themselves in regular patterns at low temperatures. Some ordering phenomena, such as crystalline, magnetic, or ferroelectric orders, have long been known and can be experimentally verified with high precision. Others, such as ferrotoroidicity\cite{Ederer2007,spaldin2008toroidal,zimmermann2014ferroic,talebi2018theory}, have remained more elusive. Composite orders are characterized by an order parameter $\left<A B\right>$, composed of two observables $A$ and $B$. \change{Interestingly, composite orders can be non-zero $\left<A B\right>\neq 0$ even if each
individual order is absent, i.e., $\left<A\right> = \left<B\right> = 0$~\cite{svistunov2015superfluid}.} Examples include odd-frequency pairing\cite{Balatsky1993,Abrahams1995,linder2019odd}, composite $U(1)$ orders, such as those composed of 2-component superconducting order parameters\cite{babaev2004superconductor, kuklov2006deconfined, babaev2004phase}, or 2-component bosonic systems\cite{kuklov2003counterflow,Kuklov2004}, and charge 4e superconductors\cite{Herland2010,Jian2021}. Recent experiments support the existence of composite orders\cite{Grinenko2021,Cho2022}. A point group symmetry approach to composite orders has been derived\cite{fernandes2019intertwined}, and composite orders have been discussed as a potential class of hidden orders\cite{aeppli2020hidden,linder2019odd}.

For ferromagnets and ferroelectrics, the order parameters are given by the magnetization $\vec{M}$ and polarization $\vec{P}$, being vectorial, i.e., multi-component order parameters. Consequently, composite order describes combinations of $\left<M_\alpha M_\beta \right>$, $\left<P_\alpha P_\beta \right>$, $\left<P_\alpha M_\beta \right>$ ($\alpha,\beta = x,y,z$).
\begin{figure}
    \centering
    \includegraphics[width=0.48\textwidth]{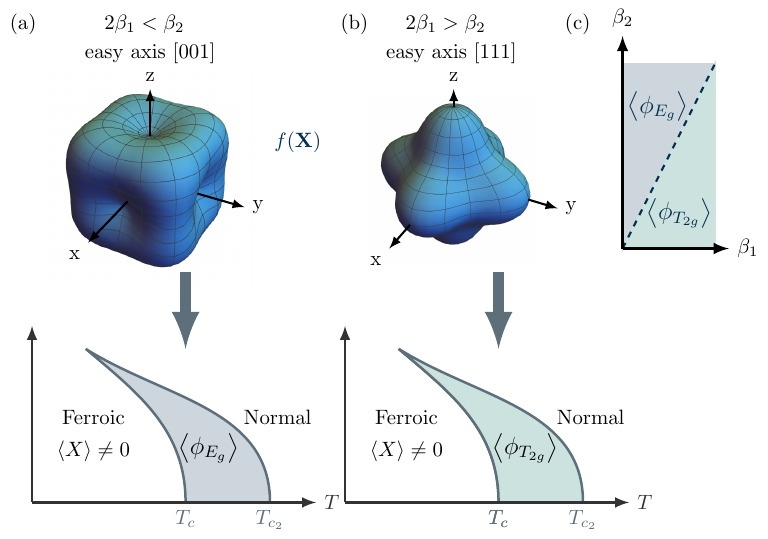}
    \caption{$Z_2$ classification of ferromagnets and ferroelectrics. Depending on the easy axis, a quadrupole order transforming as $E_g$ or $T_{2g}$ emerges above the ferroic transition temperature. The anisotropy of the cubic free energy for a vectorial order $\vec{X}$ (e.g. magnetization $\vec{M}$ or polarization $\vec{P}$) for an easy axis of (a) [100] and (b) [111]. (c) illustrates the phase diagram depending on two positive fourth order parameters $\beta_1$ and $\beta_2$.}
    \label{compositePP}
\end{figure}
We show that, depending on the anisotropy, a composite or multipolar order emerges in the vicinity of a conventional ferromagnetic or ferroelectric phase transition. For cubic crystalline symmetry, we find that cubic ferroelectrics and ferromagnets undergo a $Z_2$ classification scheme with respect to the type of quadrupole order. Here, the $Z_2$ classification is a direct consequence of the two possible easy axes for magnetization and polarization in a fourth order theory. Furthermore, we demonstrate that the concept of composite orders can be extended to multiferroic materials, where polarization and magnetization are coupled. In cubic symmetry, this coupling gives rise to three potential composite orders, aligning with the well-known multipole expansion\cite{spaldin2008toroidal,Spaldin2013,thole2018magnetoelectric}. Hence, we show that the complex phase diagram of magnetic and magnetoelectric materials is strongly dependent on composite orders, challenging the idea of competing orders.

\begin{table}[]
    \centering
    \begin{tabular}{lcc}
    \hline\hline
         & Time-reversal $\mathcal{T}$ & Inversion $\mathcal{P}$ \\
    \hline
      Polarization $\vec{P}$  & + & - \\
      Magnetization $\vec{M}$  & - & + \\
      \hline 
      Quadrupole electric $P_\alpha P_\beta$  & + & + \\
      Quadrupole magnetic $M_\alpha M_\beta$  & + & + \\
      Composite multiferroic $P_\alpha M_\beta$  & - & - \\
      \hline 
        Coupling to strain & + & + \\
        Coupling to toroidal moment & - & - \\
      \hline\hline
    \end{tabular}
    \caption{Transformation behavior under the fundamental symmetries time-reversal and inversion.}
    \label{TrP}
\end{table}
\section{Cubic ferromagnets and ferroelectrics}
The magnetization $\vec{M}$ is a pseudo-vector, which means that it is even under inversion and odd under time reversal (Table \ref{TrP}). In contrast, the polarization $\vec{P}$ behaves as an ordinary vector, being odd under inversion and even under time-reversal. In mean field theory and without lattice anisotropy, a phenomenological theory accounts only for the magnitude of $\vec{M}$ and $\vec{P}$, leading to inversion- and time-reversal-invariant free energies $f(M) = \alpha (T-T_c) M^2 + \beta M^4$ or $f(P) = \alpha (T-T_c) P^2 + \beta P^4$, with $\alpha, \beta > 0$. At high temperatures $T > T_c$, the free energy is minimized by the trivial solutions $M = 0$ and $P = 0$. Below the critical temperature $T < T_c$, a finite magnetization or polarization appears, indicating a transition into a ferromagnetic or ferroelectric state.

\change{
Beyond mean field theory, polarization and magnetization are space-dependent. Furthermore, when lattice anisotropy is considered, the free energy also depends on the direction of polarization or magnetization. In cubic symmetry, it is expressed as \cite{gtpack1,barabanov2022landau,devonshire1954theory}:
\setlength\multlinegap{0pt}
\begin{multline}
f(\vec{X},T) = f_0(T) + c\left(\nabla \vec{X}\right)^2 +  \alpha(T-T_c) \vec{X}^2 \\ + \beta_1 \left(X_x^4 + X_y^4  + X_z^4\right) + \beta_2 \left(X_x^2 X_y^2 + X_y^2 X_z^2 + X_z^2 X_x^2\right),
\label{free_energy_X}
\end{multline}
where $\vec{X} = \vec{P}, \vec{M}$ represents either the magnetization $\vec{M}$ or the polarization $\vec{P}$. In the case of antiferromagnets and antiferroelectrics, $\vec{X} = \vec{M}_1 - \vec{M}_2$ or $\vec{X} = \vec{P}_1 - \vec{P}_2$, denoting the difference in contributions from two sublattices. Furthermore, the term $f_0(T)$ accounts for the temperature-dependent free energy contributions independent of $\vec{X}$~\cite{Schwabl2006}}. Assuming second-order phase transitions and $\alpha, \beta_1, \beta_2 > 0$, we avoid higher-order terms to achieve minima at finite values $X_i$. The fourth-order terms $\beta_1$ and $\beta_2$ determine the anisotropy energy shape. In the ferroic phase ($T < T_c$), the free energy is minimized for magnetization or polarization along the Cartesian axes [100] if $2\beta_1 < \beta_2$ (Fig. \ref{compositePP}(a)). In contrast, for $2\beta_1 > \beta_2$, a magnetization or polarization along the diagonal [111] is energetically favored (Fig. \ref{compositePP}(b)). Cubic magnetic anisotropy has long been recognized \cite{Kleis1936,VanFleck1937}, with Fig. \ref{compositePP}(a) describing iron and Fig. \ref{compositePP}(b) representing nickel.

\change{
The corresponding partition function for the free energy \eqref{free_energy_X} is given by
\begin{equation}
    \mathcal{Z} = \int \mathcal{D}^3 X\, e^{- \beta \int \mathrm{d}^d x\, f(\vec{X},T)}
    \label{partitionfunction}
\end{equation}
We argue that equation \eqref{free_energy_X} and \eqref{partitionfunction} also supports a composite order beyond the standard ferroic phases. To do so, we introduce bilinears $\phi_k = \sum_{ij} c^k_{ij} X_i X_j$. Specifically, for the point group $O_h$ (full cubic symmetry), we define symmetry-adapted bilinears: $\phi_{A_{1g}}=\frac{1}{\sqrt{3}}(X_x^2+X_y^2+X_z^2)$, which transforms as the identity representation ($A_{1g}$), and the quadrupole orders $\phi_{E_{g}; 1}=\frac{1}{\sqrt{2}}(X_x^2-X_y^2)$ and $\phi_{E_{g}; 2}=\frac{1}{\sqrt{6}}(X_x^2+X_y^2-2X_z^2)$, which transform as the two-dimensional representation $E_g$. Additionally, $\vec{\phi}_{T_{2g}} = \sqrt{2} (X_x X_y,X_y X_z,X_z X_x)$ transforms as the three-dimensional irreducible representation $T_{2g}$.
}

\change{To introduce the bilinears into our theory, we use the Hubbard-Stratonovich transformation \cite{altland2010condensed}, given by
\begin{equation}
    e^{\int\mathrm{d}^d x\, \frac{a}{2} \left[\vec{X}\mathrm{M}_i \vec{X}\right]^2} = \mathcal{N} \int \mathcal{D} \phi  e^{-\int\mathrm{d}^d x\, \frac{1}{2 a} \phi^2 - \vec{X}\mathrm{M}_i \vec{X} \phi}.
\end{equation}
This allows us to write the free energy in two different forms. First, in terms of the fields $\phi_{A_{1g}}$ and $\vec{\phi}_{E_g}$, equation \eqref{free_energy_X} becomes 
\begin{multline}
    f(\vec{X},\phi_{A_{1g}},\vec{\phi}_{E_{g}},T) = f_0(T) \\ +\vec{X}\left(c \nabla^2 + r + \phi_{A_{1g}} + \mat{M}_{E_g;1} \vec{\phi}_{E_{g};1} + \mat{M}_{E_g;2} \vec{\phi}_{E_{g};2}  \right)\vec{X} \\+ \frac{1}{2\left(\beta_1+\beta_2\right) }\phi_{A_{1g}}^2 - \frac{1}{\left(2 \beta_1-\beta_2\right)} \vec{\phi}_{E_{g}}^2.
    \label{free_energy_P}
\end{multline}
Here, we absorbed the temperature-dependent second-order coefficient into $\frac{r}{2} = \alpha (T-T_c)$. The matrices $\mat{M}_{A_{1g}}$ and $\mat{M}_{E_g;i}$ are defined to satisfy $\vec{X}\cdot\mat{M}_{E_g;1}\cdot\vec{X} = \frac{1}{\sqrt{2}} \left(X_x^2-X_y^2\right) = \phi_{E_g;1}$ and $\vec{X}\cdot\mat{M}_{E_g;2}\cdot\vec{X} = \frac{1}{\sqrt{6}} \left(X_x^2+X_y^2-2 X_z^2\right)= \phi_{E_g;2}$, respectively. Furthermore, $\vec{\phi}^2_{E_g} = \phi_{E_g;1}^2 + \phi_{E_g;2}^2$.}

\change{
In terms of the fields $\phi_{A_{1g}}$ and $\vec{\phi}_{T_{2g}}$, equation \eqref{free_energy_X} can be brought into a second form, given by  
\begin{multline}
    f(\vec{X},\phi_{A_{1g}},\vec{\phi}_{T_{2g}},T) = f_0(T) \\ +\vec{X}\left(c \nabla^2 + r + \phi_{A_{1g}} + \sum_{i=1}^{3} \mat{M}_{T_{2g};i} \vec{\phi}_{T_{2g};i} \right)\vec{X} \\+ \frac{1}{2\left(\beta_1+\beta_2\right) }\phi_{A_{1g}}^2 - \frac{1}{\left(\beta_2 - 2 \beta_1\right)} \vec{\phi}_{T_{2g}}^2.
    \label{free_energy_P2}
\end{multline}
Here, we use matrices $\mat{M}_{T_{2g};i}$ according to $\vec{X}\cdot\mat{M}_{T_{2g};1}\cdot\vec{X} = \sqrt{2}X_xX_y$, $\vec{X}\cdot\mat{M}_{T_{2g};2}\cdot\vec{X} = \sqrt{2}X_yX_z$, and $\vec{X}\cdot\mat{M}_{T_{2g};3}\cdot\vec{X} = \sqrt{2}X_zX_x$.
}

\change{We show that $\phi_{A_{1g}}$, $\vec{\phi}_{E_{g}}$, and $\vec{\phi}_{T_{2g}}$ are higher rank order parameters. $\phi_{A_{1g}}$ does not break crystalline symmetries. As $\beta_1+\beta_2 > 0$, it only lowers the free energy for $T<T_c$ and is therefore not realized. In contrast, for $T>T_c$ the free energy is lowered by a non-zero value of a quadrupole order, as will be shown subsequently. A corresponding phase diagram is shown in Figure~\ref{compositePP}(c).}

\change{In the following, we focus on case 1, given in equation \eqref{free_energy_P}.} For temperatures above the ferroic transition, $T>T_c$, the fields $\vec{X}$ are fluctuating and can be integrated out, leading to the effective free energy in the fields $\phi_{A_{1g}}$ and $\vec{\phi}_{E_{g}}$ \change{(see appendix for details)},
\begin{multline}
f = \frac{1}{2\beta} \log \det\left[c\nabla^2+r + \mathrm{M}_1 \phi_{E_g;1} + \mathrm{M}_2 \phi_{E_g;2} + \phi_{A_{1g}} \right]  \\+  \frac{\phi_{A_{1g}}^2}{2 (\beta_1+\beta_2)} - \frac{\phi_{E_{g};1}^2}{2 (2\beta_1-\beta_2)} - \frac{\phi_{E_{g};2}^2}{2 (2\beta_1-\beta_2)}.
 \label{action}
 \end{multline}
The effective free energy gives rise to a self-consistent equation for the fields $\vec{\phi}_{E_{g};i}$, by evaluating the saddle point approximation $\frac{\delta f}{\delta \vec{\phi}_{E_{g};i}} = 0$. Assuming slowly varying fields $\vec{\phi}_{E_{g};i} \approx \text{const}$ one obtains a non-trivial solution if (details in the appendix)
\begin{equation}
     1 = - \frac{(2\beta_1-\beta_2) \pi}{2 k_BT \sqrt{c^3 \alpha(T-T_c)}}.
    \label{dynamic}
\end{equation}
Equation \eqref{dynamic} leads to several key conclusions. First, for the right-hand side to be positive, the condition $2\beta_1 < \beta_2$ must be satisfied. Second, the right-hand side is a real number only for $T > T_c$. Therefore, the transition into the quadrupole phase occurs at a temperature above the critical point for the dipole phase transition, i.e., the onset of ferromagnetism (antiferromagnetism) or ferroelectricity (antiferroelectricity).
\begin{figure}
    \centering
    \includegraphics[width=0.5\linewidth]{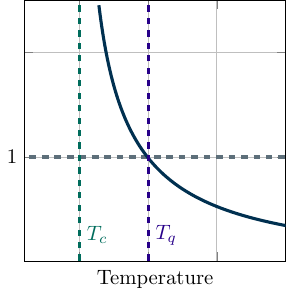}
    \caption{Evaluation of the transition temperature into the quadrupole phase. $T_c$ denotes the phase transition into the ferromagnetic or ferroelectric phase. The transition temperature to the quadrupole phase $T_q$ is determined by the blue solid line $\sim (T-Tc)^{-1/2}$ (see equation \eqref{dynamic}) being equal to 1. }
    \label{fig:tc}
\end{figure}

\change{More specifically, equation \eqref{dynamic} is shown in Figure \ref{fig:tc}. For $2\beta_1 < \beta_2$, the right hand side of \eqref{dynamic} is a positive and monotonically decreasing function in the interval $T>T_c$. In particular, it has a singularity at $T=T_c$ and vanishes for $T\rightarrow \infty$. Hence, it always passes through $1$ indicating the phase transition into the quadrupole phase. 
%
%
%
The difference between the critical temperatures for the dipole phase $T_c$ and the quadrupole phase $T_q$ depends on the anisotropy $2\beta_1 - \beta_2$. Hence, for weakly anisotropic materials $2 \beta_1 \approx \beta_2$ the transition temperatures $T_q$ and $T_c$ coincide and the phase transition is not observed. This indicates, why a quadrupole transition has not been observed in most simple ferromagnets.}

While there is no microscopic theory yet for the phenomenological theory developed in this paper, several materials have been verified with a quadrupole tranistion occuring in the vicinity of a dipole transition. For magnetic materials, magnetic anisotropy is driven by strong spin-orbit interaction. An example for a strong spin-orbit antiferromagnet are rare earth hexaborites, with e.g., DyB$_6$ having a Nèel temperature of $T_c \approx 25$~K and a quadrupole transition at $T_q \approx 30.2~\text{K}$~\cite{Goto2000}. 
\begin{figure*}
    \centering
    \includegraphics[width=0.98\textwidth]{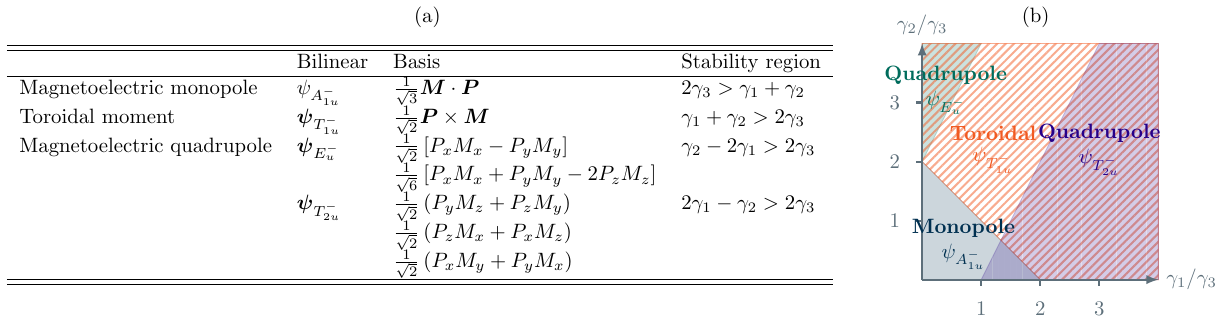}
    \caption{Composite multiferroic orders in cubic symmetry. (a) Bilinears for composite multiferroic orders and stability region. (b) Phase diagram of composite multiferroic orders.}
    \label{tab:MF_cubic}
\end{figure*}

Another promising example is the cubic canted antiferromagnet Ba$_2$MgReO$_6$. According to magnetization, muon spin relaxation, and neutron-scattering performed by Marjerrison \textit{et al.}, a magnetic phase transition occurs below 18~K, accompanied by a quadrupolar phase between 18~K and $\approx 33~$K \cite{marjerrison2016cubic}. Both phase transitions can clearly be seen by peaks in the specific heat. The phase diagram could be verified by synchrotron x-ray-diffraction measurements, reported by Hirai \textit{et al.} \cite{Hirai2020}. Ba$_2$MgReO$_6$ has been investigated more extensively by ab initio methods confirming the strong influence of spin-orbit interaction in the formation of the quadrupole order \cite{pasztorova2023experimental,ketfi2024mechanical,Tehrani2021}. Similar behavior is also observed in Ba$_2$ZnReO$_6$ \cite{marjerrison2016cubic}, having a dipole order at $T_c \approx 11~\text{K}$ and a quadrupole order at $\approx 33~\text{K}$.

Additionally, the cubic spinel selenide GaNb$_4$Se$_8$ shows a quadrupole phase with a transition temperature of $T_q = 50~\text{K}$ and an antiferromagnetic transition at $T_c = 33~\text{K}$ \cite{Ishikawa2020}. The microscopic theory for this behavior accounts for a combination of strong correlations and spin-orbit interaction. 

The composite orders $\vec{\phi}_{E_{g}}$ and $\vec{\phi}_{T_{2g}}$ break the cubic symmetry and are even under inversion and time-reversal, regardless of its origin (magnetization or polarization), see Table~\ref{TrP}. Hence, they neither couple linearly to an external electric nor magnetic field. Instead, the electric or magnetic fields couple quadratically to their respective order parameters, which would be visible in a modification of the fourth order susceptibility, as recently discussed on the example of the hidden order phase in URu$_2$Si$_2$ \cite{Trinh2016}. However, $\vec{\phi}_{E_{g}}$ and $\vec{\phi}_{T_{2g}}$ couple linearly to the respective components of the strain tensor, \cite{Brassington1982}. Hence, elastic anomalies prior to the phase transition can be expected.

Such behavior has been verified in ferroelectrics. For example, in PbSc$_{0.5}$Ta$_{0.5}$O$_3$ a precursor regime was identified ranging over $\approx 100~K$ above the ferroelectric transition, exhibiting a softening of the shear elastic constant, which was initially explained by a higher order coupling of coexisting order parameters \cite{Aktas2013}. In contrast, the composite ferroelectric phase couples linearly to the shear strain, potentially explaining the strength of the effect.

\section{Multiferroics}
\change{We continue by extending the model of composite orders to multiferroics \cite{fiebig2005revival,KHOMSKII20061,Ramesh2007,Fiebig2016,Tokura_2014,spaldin2019,bossini2023magnetoelectrics}, i.e., materials with coexisting ferroelectric and magnetic ordering.} As before, we focus on materials where the high temperature phase (before the ferroelectric and magnetic phase transition) is cubic. Such a case is realized in the multiferroic perovskites, e.g. BiFeO$_3$ \cite{Wang2003}, which has a cubic crystal structure at temperatures above $\approx1,103$~K~ \cite{Lee2015}. The free energy of a multiferroic contains a contribution from the ferroelectric phase $f(\vec{P})$, the magnetic phase $f(\vec{M})$, and their coupling $f_c$,\change{
\begin{equation}
    f(\vec{P},\vec{M},T) = f(\vec{P},T) + f(\vec{M},T) + f_c.
\end{equation}}
In a time-reversal invariant system with cubic symmetry, the coupling term contains three linearly independent contributions, given by
\begin{multline}
    f_c = \gamma_1 \left(P_x^2M_x^2+P_y^2M_y^2+P_z^2M_z^2\right) \\ + \gamma_2 \left(P_xP_yM_xM_y + P_yP_zM_yM_z + P_zP_xM_zM_x\right) \\
    +\gamma_3 \left(P_x^2(M_y^2 + M_z^2) + P_y^2(M_z^2 + M_x^2) \right. \\+ \left. P_z^2(M_x^2 + M_y^2)\right).
    \label{fc:MF}
\end{multline}
As the formation of composite orders in the magnetic and ferroelectric degrees of freedom has been discussed before, we now focus on the potential of forming multiferroic composites $\left<P_\alpha M_\beta\right>$, emerging from the multiferroic coupling $f_c$. As before, we decompose $f_c$ into symmetry adapted bilinears $\psi_k = \sum_{ij} c_{ij}^k P_i M_j$. Due to the individual symmetries of polarization and magnetization, the bilenars $\psi_k$ are odd under time-reversal $\mathcal{T}$ and odd under inversion $\mathcal{P}$. 

In cubic symmetry, the tensor product of the polarization and the magnetization gives $T_{1u}^+ \otimes T_{1g}^- \simeq A_{1u}^- \oplus T_{1u}^- \oplus E_u^- \oplus T_{2u}^-$. These representations can be interpreted in terms of the magnetoelectric multipole expansion \cite{spaldin2008toroidal,Spaldin2013}, with $A_{1u}^-$ being the pseudoscalar describing the magnetoelectric monopole, $T_{1u}^-$ the toroidal moment vector, and $E_u^-$ and $T_{2u}^-$ the tensor describing the quadrupole magnetic moment. The corresponding symmetry adapted bilinears in $\vec{P}$ and $\vec{M}$ are given in Figure~\ref{tab:MF_cubic}(a). As the tensor product of polarization and magnetization can be decomposed into terms transforming as 4 irreducible representations ($A_{1u}^-, T_{1u}^-, E_u^-, T_{2u}^-$), decomposing the fourth order terms should give rise to four scalars symmetric under all cubic symmetries, $\left(T_{1u}^+ \otimes T_{1g}^-\right)^2 \simeq 4 A_{1g} \oplus \dots$ However, as the components of $\vec{P}$ and $\vec{M}$ commute, only 3 linearly independent contributions remain. This becomes apparent from the free energy in equation \eqref{fc:MF}. As a result, the free energy in \eqref{fc:MF} can be expressed in three, instead of four bilinears (comparable to the Fierz identity in high energy physics). For example, eliminating $\phi_{A_{1u}}^-$ and writing \eqref{fc:MF} in terms of $\vec{\psi}_{E_{u}^-}$, $\vec{\psi}_{T_{1u}^-}$, and $\vec{\psi}_{T_{2u}^-}$ the multiferroic coupling $f_c$ is expressed as follows,
\begin{equation}
    f_c = \frac{3\gamma_1}{2} \vec{\psi}_{E_{u}^-}^2 + \frac{2\gamma_3-\gamma_1-\gamma_2}{2}\vec{\psi}_{T_{1u}^-}^2 + \frac{2\gamma_3+\gamma_1+\gamma_2}{2} \vec{\psi}_{T_{2u}^-}^2.
    \label{fc}
\end{equation}
As before, we obtain a free energy with second order contributions in the bilinears, which serve as order parameters of the three corresponding types of composite multiferroic order. Assuming that a fourth order theory is sufficient to describe a system in mind, we require $\gamma_1,\gamma_2,\gamma_3 > 0$. Hence, while the coefficients before the second order terms in $\vec{\psi}_{E_{u}^-}$ and $\vec{\psi}_{T_{2u}^-}$ are always possitive, the toroidal moment $\vec{\psi}_{T_{1u}^-}$ can emerge as a composite order parameter if $\gamma_1 + \gamma_2 > 2 \gamma_3$. Following the theory of composite orders, a macroscopic toroidal moment in the sample \cite{Ederer2007,spaldin2008toroidal,talebi2018theory}, can be found as a high temperature phase in a strongly coupled multiferroic material, or at low temperatures in incipient multiferroics. By symmetry, the toroidal order couples to the curl of a magnetic field, $\sim \vec{\psi}_{T_{1u}^-} \cdot \nabla\times\vec{B}$. Furthermore, the toroidal order induces a magnetoelectric effect, as can be seen from the polarizability $\vec{P} = \chi^e \epsilon_0 \vec{E} + \eta \vec{B}\times \vec{\psi}_{T_{1u}^-}$ \cite{Ederer2007}.

In a similar logic as for the toroidal order, different versions of the free energy can be derived to stabilize the monopole or quadrupole orders. For example, by eliminating $\phi_{T_{1u}^-}$ the free energy \eqref{fc:MF} can be written in terms of $\psi_{A_{1u}^-}$, $\vec{\phi}_{E_{u}^-}$, and $\vec{\phi}_{T_{2u}^-}$,
\begin{equation}
    f_c =  \left(\gamma_1+\gamma_2-2\gamma_3\right)\psi_{A_{1u}^-}^2 + \frac{2\gamma_1-\gamma_2+2\gamma_3}{2} \vec{\psi}_{E_{u}^-}^2 + 2\gamma_3 \vec{\psi}_{T_{2u}^-}^2.
    \label{fc2}
\end{equation}
As a result, for  $\gamma_1,\gamma_2,\gamma_3 > 0$, the free energy \eqref{fc2} gives rise to the magnetoelectric monopole phase for $2\gamma_3 > \gamma_1 + \gamma_2$ and the magnetic quadrupole phase for $\gamma_2 > 2(\gamma_1+\gamma_3)$. 

The phase diagram for all four possible composite phases is summarized in Figure~\ref{tab:MF_cubic}(b). Interestingly, while there are regions where the monopole or toroidal order can be established independently, the stability regions for the quadrupole orders are found to compete with the toroidal and monopole orders. Under which conditions these orders can be established would require a separate discussion beyond the scope of this paper.

\section{Summary and outlook}
In summary, we showed the existence of composite orders in ferromagnetic, ferroelectric, and multiferroic materials \change{(the approach works similarly for antiferromagnets and antiferroelectrics, respectively)}. Starting with cubic ferromagnets and ferroelectrics, we found that a fourth-order Landau expansion of the order parameter gives rise to a quadrupole order for sufficiently strong cubic anisotropy. For materials with an easy axis along the [100] direction, the quadrupole order exhibits a $E_g$ symmetry while for an easy axis along the [111] axis, a quadrupole order with $T_{2g}$ symmetry is observed. More explicitly, we showed that the quadrupole transition temperature occurs above the dipole transition temperature for an ferromagnetic or ferroelectric order, respectively. Using a field theoretical perspective, we find that the difference between dipole and quadrupole transition temperatures depends on the anisotropy energy. As a result it vanishes for most simple ferromagnets and ferroelectrics. The theory of composite orders derived here is in qualitative agreement with reports about quadrupole orders above magnetic phase transitions, e.g., in Ba$_2$MgReO$_6$ \cite{marjerrison2016cubic,Hirai2020}, Ba$_2$ZnReO$_6$ \cite{marjerrison2016cubic}, GaNb$_4$Se$_8$ \cite{Ishikawa2020}, and RB$_6$ (R = H, Dy) \cite{Goto2000}. By symmetry, the $E_g$ and $T_{2g}$ quadrupole orders couple to strain applied to the respective material.  

In the second part, we extended the discussion to multiferroics, using the specific example of a coexisting polarization and magnetization in a cubic system. While we primarily focussed on ferromagnetic order, a similar argument holds for antiferromagnetic or more complex magnetic orders. Incorporating a fourth order coupling term in the Landau theory, we showed that the coupling term can be decomposed into four different bilinears describing electromagnetic composite order, being the magnetoelectric monopole ($A_{1u}^-$), the toroidal order ($T_{1u}^-$) as well as magnetoelectric quadrupole ordfers ($T_{2u}^-$ and $E_u^-$). In terms of the corresponding composite order parameters, the free energy can be formulated in four different variants, giving rise to regions where these orders can emerge at elevated temperatures.

In summary, we could show that the theory of composite orders, can be extended to ferroic materials. Our result shows that the complex phase diagrams in ferroic, multiferroic, and magentoelectric materials cannot be seen as composed of individual competing orders, but instead as strongly dependent phase transitions. Furthermore, this interpretation becomes interesting when the primary order is not seen at temperature above 0~K, e.g., in incipient ferroelectrics. Still, a composite order above 0~K is allowed. This argument is relevant for the ongoing discussion of hidden order in the condensed matter. For example, in URu$_{2-x}$Fe$_x$Si$_2$ a clear phase transition occurs at temperatures above an antiferromagnetic order, in the region around $x\approx 0.1$ \cite{Kung2016}.

\begin{acknowledgments}
We acknowledge inspiring discussions with Alexander Balatsky, Nicola Spaldin, Egor Babaev, Hennadii Yerzhakov, and Wolfram Hergert. We are grateful for funding from the Swedish Research Council (VR starting Grant No. 2022-03350), the Olle Engkvist Foundation (Grant No. 229-0443), the Royal Physiographic Society in Lund (Horisont), the Knut and Alice Wallenberg Foundation (Grant No. 2023.0087), and Chalmers University of Technology, via the department of physics and the Areas of Advance Nano and Materials.
\end{acknowledgments}

\section*{Appendix}

\subsection*{Decomposing the multiferroic coupling in symmetry-adapted bilinears}
The multiferroic coupling between polarization $\vec{P}$ and magnetization $\vec{M}$ for cubic symmetry is given in the main text, equation \eqref{fc:MF}. As it contains three linearly independent fourth order terms, four variants can be derived in terms of a decomposition into the four allowed symmetry adapted bilinears  $\psi_{A_{1u}^-}$, $\vec{\phi}_{E_{u}^-}$, $\vec{\phi}_{T_{1u}^-}$ and $\vec{\phi}_{T_{2u}^-}$. We provide these expressions subsequently. Eliminating $\phi_{A_{1u}}^-$, the multiferroic coupling in the free energy is given by 
\begin{equation}
    f_c^{(1)} = \frac{3\gamma_1}{2} \vec{\psi}_{E_{u}^-}^2 + \frac{2\gamma_3-\gamma_1-\gamma_2}{2}\vec{\psi}_{T_{1u}^-}^2 + \frac{2\gamma_3+\gamma_1+\gamma_2}{2} \vec{\psi}_{T_{2u}^-}^2.
\end{equation}
Eliminating $\vec{\phi}_{E_{u}}^-$, one obtains  
\begin{equation}
    f_c^{(2)} = 3\gamma_1 \psi_{A_{1u}^-}^2 + \frac{2\gamma_1-\gamma_2+2\gamma_3}{2}\vec{\psi}_{T_{1u}^-}^2 + \frac{2\gamma_3-2\gamma_1+\gamma_2}{2} \vec{\psi}_{T_{2u}^-}^2.
\end{equation}
Eliminating $\vec{\phi}_{T_{1u}}^-$, one obtains  
\begin{equation}
    f_c^{(3)} =  \left(\gamma_1+\gamma_2-2\gamma_3\right)\psi_{A_{1u}^-}^2 + \frac{2\gamma_1-\gamma_2+2\gamma_3}{2} \vec{\psi}_{E_{u}^-}^2 + 2\gamma_3 \vec{\psi}_{T_{2u}^-}^2.
\end{equation}
Eliminating $\vec{\phi}_{T_{2u}}^-$, one obtains  
\begin{equation}
    f_c^{(4)} =  \left(\gamma_1+\gamma_2+2\gamma_3\right)\psi_{A_{1u}^-}^2 + \frac{2\gamma_1-\gamma_2-\gamma_3}{2} \vec{\psi}_{E_{u}^-}^2 + 2\gamma_3 \vec{\psi}_{T_{1u}^-}^2.
\end{equation}

\subsection*{Field theory treatment of composite order}
We promote the vector $\vec{X}$ to a field $\vec{X}(\vec{r},t)$. We add a gradient term $\frac{\vec{X} \nabla^2 \vec{X}}{2}$ to the free energy \eqref{free_energy_X} and bring it into the form of equation \eqref{free_energy_P}, 
\begin{multline}
    f(\vec{X},T) = f_0(T) + \frac{c}{2}\vec{X} \nabla^2 \vec{X} + \frac{r}{2} \vec{X}^2 + \left(\beta_1 + \beta_2 \right) \left(\vec{X}^2\right)^2 \\ + \left(\frac{2\beta_1-\beta_2}{2}\right)\left(\left[\vec{X}\mathrm{M}_1 \vec{X}\right]^2 + \left[\vec{X}\mathrm{M}_2 \vec{X}\right]^2\right), 
\end{multline}
with $c$ being a constant. For brevity, we introduced $\frac{r}{2} = \alpha \left(T - T_c\right)$. The matrices $\mathrm{M}_1$ and $\mathrm{M}_2$ are chosen to satisfy, $\vec{X}\mathrm{M}_1\vec{X} = \frac{1}{\sqrt{2}} \left(X_x^2-X_y^2\right) = \phi_{E_g; 1}$ and $\vec{X}\mathrm{M}_2\vec{X} = \frac{1}{\sqrt{6}} \left(X_x^2+X_y^2-2 X_z^2\right) = \phi_{E_g; 2}$, i.e.,
\begin{equation}
    \mathrm{M}_1 = \frac{1}{\sqrt{2}}\left(\begin{array}{ccc}
        1 & 0 & 0 \\
        0 & -1 & 0 \\
        0 & 0 & 0 
    \end{array}\right), \quad 
    \mathrm{M}_2 = \frac{1}{\sqrt{6}}\left(\begin{array}{ccc}
        1 & 0 & 0 \\
        0 & 1 & 0 \\
        0 & 0 & -2 
    \end{array}\right).
\end{equation}
The corresponding partition function is given by
\begin{equation}
    \mathcal{Z} = \int \mathcal{D}^3 X\, e^{- \beta \int \mathrm{d}^d x\, f(\vec{X})},
\end{equation}
with $\beta = 1/k_B T$. To introduce the quadrupolar order, we apply the Hubbard-Stratonovich transformation. This step is based on the Gaussian integration and the equation,
\begin{equation}
    e^{\int\mathrm{d}^d x\, \frac{a}{2} \left[\vec{X}\mathrm{M}_i \vec{X}\right]^2} = \mathcal{N} \int \mathcal{D} \phi  e^{-\int\mathrm{d}^d x\, \frac{1}{2 a} \phi^2 - \vec{X}\mathrm{M}_i \vec{X} \phi},
\end{equation}
with $\mathcal{N}$ being a normalization. Identifying $a = - (2\beta_1 - \beta_2)$ for the fields $\phi_{E_{g};i}$ and $a = \beta_1 + \beta_2$ for the field $\phi_{A_{1g}}$ leads to the following expression of the partition function,
\begin{multline}
    \mathcal{Z} = \mathcal{N} \mathcal{Z}_0 \int \mathcal{D}^3X \,\mathcal{D}\phi_{A_{1g}} \,\mathcal{D}\phi_{E_{g};1} \,\mathcal{D}\phi_{E_{g};2}\\ \exp \left[ - \beta \int \mathrm{d}^d x  
    \vec{X} \left(\frac{c\nabla^2}{2}+\frac{r}{2}\right)\vec{X} + \vec{X}^2 \phi_{A_{1g}}  \right.\\ \left.
    + \frac{\phi_{A_{1g}}^2}{2 (\beta_1+\beta_2)} + \vec{X}\mathrm{M}_1 \vec{X} \phi_{E_g;1} + \vec{X}\mathrm{M}_2 \vec{X} \phi_{E_g;2} \right. \\ \left. - \frac{\phi_{E_{g};1}^2}{2 (2\beta_1-\beta_2)} - \frac{\phi_{E_{g};2}^2}{2 (2\beta_1-\beta_2)}
    \right].
\end{multline}
\change{Here, we absorbed the $\vec{X}$-independent part in the free energy, $f_0(T)$ into the term $\mathcal{Z}_0$.} Above the phase transition into the ferroelectric or ferromagnetic state, $T>T_c$, $\vec{X}$ is fluctuating and can be integrated out. As a result, one obtains
\begin{multline}
    \mathcal{Z} = \mathcal{N} \mathcal{Z}_0\int \mathcal{D}\phi_{A_{1g}} \,\mathcal{D}\phi_{E_{g};1} \,\mathcal{D}\phi_{E_{g};2} \\ \det\left[c\nabla^2+r + \mathrm{M}_1 \phi_{E_g;1} + \mathrm{M}_2 \phi_{E_g;2} + \phi_{A_{1g}} \right]^{-\frac{1}{2}} \\ \\ \exp \left[ - \beta \int \mathrm{d}^d x \,
    \frac{\phi_{A_{1g}}^2}{2 (\beta_1+\beta_2)}  -\frac{\phi_{E_{g};1}^2}{2 (2\beta_1-\beta_2)} - \frac{\phi_{E_{g};2}^2}{2 (2\beta_1-\beta_2)}
    \right].
\end{multline}
This gives rise to the free energy, $f = -k_B T \ln \mathcal{Z}$, 
\begin{multline}
 f = \frac{1}{2 \beta} \ln \det\left[c\nabla^2+r + \mathrm{M}_1 \phi_{E_g;1} + \mathrm{M}_2 \phi_{E_g;2} + \phi_{A_{1g}} \right]  \\+  \frac{\phi_{A_{1g}}^2}{2 (\beta_1+\beta_2)} - \frac{\phi_{E_{g};1}^2}{2 (2\beta_1-\beta_2)} - \frac{\phi_{E_{g};2}^2}{2 (2\beta_1-\beta_2)}.
 \label{free_energy}
\end{multline}
This free energy allows us to determine a gap equation for the composite orders $\phi_{E_g;i}$. We use the saddle point approximation 
\begin{equation}
 \frac{\delta f[\phi_{E_g;i}]}{\delta \phi_{E_g;i}} = 0.   
 \label{saddle}
\end{equation}
To evaluate the saddle point equation, we use the identity: $\ln \det \hat{\mathrm{G}_i}^{-1} = \operatorname{tr} \ln \hat{\mathrm{G}_i}^{-1}$ \cite{Altland2010} and set $\hat{\mathrm{G}_i}^{-1} = c\nabla^2+r + \mathrm{M}_i \phi_{E_g;i}$. We now derive \cite{Altland2010},
\begin{equation}
  \frac{\delta}{\delta \phi_{E_g;i}}  \operatorname{tr} \ln \hat{\mathrm{G}_i}^{-1} = \operatorname{tr}\left[\hat{\mathrm{G}_i} \frac{\delta}{\delta \phi_{E_g;i}} \hat{\mathrm{G}_i}^{-1}\right].
\end{equation}
We evaluate the trace, both in matrix space and in operator space ($\operatorname{tr} \hat{\mathrm{A}} = \sum_{\vec{k}} \left< \vec{k}\right|\hat{\mathrm{A}} \left| \vec{k}\right>$). We obtain for the two components of the quadrupole order,
\begin{equation}
 \operatorname{tr}\left[\hat{\mathrm{G}_1} \frac{\delta \hat{\mathrm{G}_1}^{-1}}{\delta \phi_{E_g;1}} \right] = \int \frac{\mathrm{d}^dk}{(2\pi)^3}\,\frac{2 \phi_{E_g;1}}{-2 (ck^2+r)^2+\phi_{E_g;1}}, \label{phi1}
\end{equation}
\begin{multline}
 \operatorname{tr}\left[\hat{\mathrm{G}_2} \frac{\delta \hat{\mathrm{G}_2}^{-1}}{\delta \phi_{E_g;2}} \right] =   \int \frac{\mathrm{d}^dk}{(2\pi)^3}\,\\\frac{18 \phi_{E_g;2}}{(-3 ck^2+ -3r + \sqrt{6}\phi_{E_g;2}) (6 ck^2 + 6r + \sqrt{6}\phi_{E_g;2})}.\label{phi2}
\end{multline}
Close to the phase transition into the quadrupole phase we assume $\phi_{E_g;i} \ll k^4$ for most momenta $k$ in the integral. Hence, we remove $\phi_{E_{g};i}$ from the denominator in equations \eqref{phi1} and \eqref{phi2} and obtain the universal equation,
\begin{equation}
   \operatorname{tr}\left[\hat{\mathrm{G}_i} \frac{\delta \hat{\mathrm{G}_i}^{-1}}{\delta \phi_{E_g;i}} \right] =  - \int \frac{\mathrm{d}^dk}{(2\pi)^3} \frac{ \phi_{E_g;i}}{\left(ck^2 + r\right)^2}.\label{universal}
\end{equation}
The corresponding linearized gap equation follows from equations \eqref{saddle}, \eqref{free_energy}, and \eqref{universal},
\begin{equation}
    \phi_{E_{g};i} = - \frac{2 (2\beta_1-\beta_2)}{k_B T} \int \frac{\mathrm{d}^dk}{(2\pi)^3}\, \frac{\phi_{E_{g};i}}{\left(ck^2+r\right)^2}. \label{gap_integral}
\end{equation}
To evaluate \eqref{gap_integral} we assume $\phi_{E_{g};i} \approx \text{constant}$. Hence, $\phi_{E_{g};i}$ cancels on both sides of \eqref{gap_integral} and the integration of $k$ on the right hand side can be performed to obtain 
\begin{equation}
    1 = -\frac{(2\beta_1-\beta_2) \pi}{2k_B T \sqrt{c^3 r}} = - \frac{(2\beta_1-\beta_2) \pi}{2 k_BT \sqrt{c^3 \alpha(T-T_c)}}.
\end{equation}

\begin{thebibliography}{49}%
\makeatletter
\providecommand \@ifxundefined [1]{%
 \@ifx{#1\undefined}
}%
\providecommand \@ifnum [1]{%
 \ifnum #1\expandafter \@firstoftwo
 \else \expandafter \@secondoftwo
 \fi
}%
\providecommand \@ifx [1]{%
 \ifx #1\expandafter \@firstoftwo
 \else \expandafter \@secondoftwo
 \fi
}%
\providecommand \natexlab [1]{#1}%
\providecommand \enquote  [1]{``#1''}%
\providecommand \bibnamefont  [1]{#1}%
\providecommand \bibfnamefont [1]{#1}%
\providecommand \citenamefont [1]{#1}%
\providecommand \href@noop [0]{\@secondoftwo}%
\providecommand \href [0]{\begingroup \@sanitize@url \@href}%
\providecommand \@href[1]{\@@startlink{#1}\@@href}%
\providecommand \@@href[1]{\endgroup#1\@@endlink}%
\providecommand \@sanitize@url [0]{\catcode `\\12\catcode `\$12\catcode
  `\&12\catcode `\#12\catcode `\^12\catcode `\_12\catcode `\%12\relax}%
\providecommand \@@startlink[1]{}%
\providecommand \@@endlink[0]{}%
\providecommand \url  [0]{\begingroup\@sanitize@url \@url }%
\providecommand \@url [1]{\endgroup\@href {#1}{\urlprefix }}%
\providecommand \urlprefix  [0]{URL }%
\providecommand \Eprint [0]{\href }%
\providecommand \doibase [0]{https://doi.org/}%
\providecommand \selectlanguage [0]{\@gobble}%
\providecommand \bibinfo  [0]{\@secondoftwo}%
\providecommand \bibfield  [0]{\@secondoftwo}%
\providecommand \translation [1]{[#1]}%
\providecommand \BibitemOpen [0]{}%
\providecommand \bibitemStop [0]{}%
\providecommand \bibitemNoStop [0]{.\EOS\space}%
\providecommand \EOS [0]{\spacefactor3000\relax}%
\providecommand \BibitemShut  [1]{\csname bibitem#1\endcsname}%
\let\auto@bib@innerbib\@empty
\bibitem [{\citenamefont {Ederer}\ and\ \citenamefont
  {Spaldin}(2007)}]{Ederer2007}%
  \BibitemOpen
  \bibfield  {author} {\bibinfo {author} {\bibfnamefont {C.}~\bibnamefont
  {Ederer}}\ and\ \bibinfo {author} {\bibfnamefont {N.~A.}\ \bibnamefont
  {Spaldin}},\ }\bibfield  {title} {\bibinfo {title} {Towards a microscopic
  theory of toroidal moments in bulk periodic crystals},\ }\href
  {https://doi.org/10.1103/PhysRevB.76.214404} {\bibfield  {journal} {\bibinfo
  {journal} {Physical Review B}\ }\textbf {\bibinfo {volume} {76}},\ \bibinfo
  {pages} {214404} (\bibinfo {year} {2007})}\BibitemShut {NoStop}%
\bibitem [{\citenamefont {Spaldin}\ \emph {et~al.}(2008)\citenamefont
  {Spaldin}, \citenamefont {Fiebig},\ and\ \citenamefont
  {Mostovoy}}]{spaldin2008toroidal}%
  \BibitemOpen
  \bibfield  {author} {\bibinfo {author} {\bibfnamefont {N.~A.}\ \bibnamefont
  {Spaldin}}, \bibinfo {author} {\bibfnamefont {M.}~\bibnamefont {Fiebig}},\
  and\ \bibinfo {author} {\bibfnamefont {M.}~\bibnamefont {Mostovoy}},\
  }\bibfield  {title} {\bibinfo {title} {The toroidal moment in
  condensed-matter physics and its relation to the magnetoelectric effect},\
  }\href {https://doi.org/10.1088/0953-8984/20/43/434203} {\bibfield  {journal}
  {\bibinfo  {journal} {Journal of Physics: Condensed Matter}\ }\textbf
  {\bibinfo {volume} {20}},\ \bibinfo {pages} {434203} (\bibinfo {year}
  {2008})}\BibitemShut {NoStop}%
\bibitem [{\citenamefont {Zimmermann}\ \emph {et~al.}(2014)\citenamefont
  {Zimmermann}, \citenamefont {Meier},\ and\ \citenamefont
  {Fiebig}}]{zimmermann2014ferroic}%
  \BibitemOpen
  \bibfield  {author} {\bibinfo {author} {\bibfnamefont {A.~S.}\ \bibnamefont
  {Zimmermann}}, \bibinfo {author} {\bibfnamefont {D.}~\bibnamefont {Meier}},\
  and\ \bibinfo {author} {\bibfnamefont {M.}~\bibnamefont {Fiebig}},\
  }\bibfield  {title} {\bibinfo {title} {Ferroic nature of magnetic toroidal
  order},\ }\href {https://doi.org/10.1038/ncomms5796} {\bibfield  {journal}
  {\bibinfo  {journal} {Nature communications}\ }\textbf {\bibinfo {volume}
  {5}},\ \bibinfo {pages} {4796} (\bibinfo {year} {2014})}\BibitemShut
  {NoStop}%
\bibitem [{\citenamefont {Talebi}\ \emph {et~al.}(2018)\citenamefont {Talebi},
  \citenamefont {Guo},\ and\ \citenamefont {van Aken}}]{talebi2018theory}%
  \BibitemOpen
  \bibfield  {author} {\bibinfo {author} {\bibfnamefont {N.}~\bibnamefont
  {Talebi}}, \bibinfo {author} {\bibfnamefont {S.}~\bibnamefont {Guo}},\ and\
  \bibinfo {author} {\bibfnamefont {P.~A.}\ \bibnamefont {van Aken}},\
  }\bibfield  {title} {\bibinfo {title} {Theory and applications of toroidal
  moments in electrodynamics: their emergence, characteristics, and
  technological relevance},\ }\href {https://doi.org/10.1515/nanoph-2017-0017}
  {\bibfield  {journal} {\bibinfo  {journal} {Nanophotonics}\ }\textbf
  {\bibinfo {volume} {7}},\ \bibinfo {pages} {93} (\bibinfo {year}
  {2018})}\BibitemShut {NoStop}%
\bibitem [{\citenamefont {Svistunov}\ \emph {et~al.}(2015)\citenamefont
  {Svistunov}, \citenamefont {Babaev},\ and\ \citenamefont
  {Prokof'ev}}]{svistunov2015superfluid}%
  \BibitemOpen
  \bibfield  {author} {\bibinfo {author} {\bibfnamefont {B.~V.}\ \bibnamefont
  {Svistunov}}, \bibinfo {author} {\bibfnamefont {E.~S.}\ \bibnamefont
  {Babaev}},\ and\ \bibinfo {author} {\bibfnamefont {N.~V.}\ \bibnamefont
  {Prokof'ev}},\ }\href {https://doi.org/10.1201/b18346} {\emph {\bibinfo
  {title} {Superfluid states of matter}}}\ (\bibinfo  {publisher} {CRC Press,
  Boca Raton},\ \bibinfo {year} {2015})\BibitemShut {NoStop}%
\bibitem [{\citenamefont {Balatsky}\ and\ \citenamefont
  {Bonc\v{a}}(1993)}]{Balatsky1993}%
  \BibitemOpen
  \bibfield  {author} {\bibinfo {author} {\bibfnamefont {A.~V.}\ \bibnamefont
  {Balatsky}}\ and\ \bibinfo {author} {\bibfnamefont {J.}~\bibnamefont
  {Bonc\v{a}}},\ }\bibfield  {title} {\bibinfo {title} {Even- and odd-frequency
  pairing correlations in the one-dimensional {$t-J-h$} model: A comparative
  study},\ }\href {https://doi.org/10.1103/PhysRevB.48.7445} {\bibfield
  {journal} {\bibinfo  {journal} {Physical Review B}\ }\textbf {\bibinfo
  {volume} {48}},\ \bibinfo {pages} {7445} (\bibinfo {year}
  {1993})}\BibitemShut {NoStop}%
\bibitem [{\citenamefont {Abrahams}\ \emph {et~al.}(1995)\citenamefont
  {Abrahams}, \citenamefont {Balatsky}, \citenamefont {Scalapino},\ and\
  \citenamefont {Schrieffer}}]{Abrahams1995}%
  \BibitemOpen
  \bibfield  {author} {\bibinfo {author} {\bibfnamefont {E.}~\bibnamefont
  {Abrahams}}, \bibinfo {author} {\bibfnamefont {A.}~\bibnamefont {Balatsky}},
  \bibinfo {author} {\bibfnamefont {D.~J.}\ \bibnamefont {Scalapino}},\ and\
  \bibinfo {author} {\bibfnamefont {J.~R.}\ \bibnamefont {Schrieffer}},\
  }\bibfield  {title} {\bibinfo {title} {Properties of odd-gap
  superconductors},\ }\href {https://doi.org/10.1103/PhysRevB.52.1271}
  {\bibfield  {journal} {\bibinfo  {journal} {Physical Review B}\ }\textbf
  {\bibinfo {volume} {52}},\ \bibinfo {pages} {1271} (\bibinfo {year}
  {1995})}\BibitemShut {NoStop}%
\bibitem [{\citenamefont {Linder}\ and\ \citenamefont
  {Balatsky}(2019)}]{linder2019odd}%
  \BibitemOpen
  \bibfield  {author} {\bibinfo {author} {\bibfnamefont {J.}~\bibnamefont
  {Linder}}\ and\ \bibinfo {author} {\bibfnamefont {A.~V.}\ \bibnamefont
  {Balatsky}},\ }\bibfield  {title} {\bibinfo {title} {Odd-frequency
  superconductivity},\ }\href@noop {} {\bibfield  {journal} {\bibinfo
  {journal} {Reviews of Modern Physics}\ }\textbf {\bibinfo {volume} {91}},\
  \bibinfo {pages} {045005} (\bibinfo {year} {2019})}\BibitemShut {NoStop}%
\bibitem [{\citenamefont {Babaev}\ \emph {et~al.}(2004)\citenamefont {Babaev},
  \citenamefont {Sudb{\o}},\ and\ \citenamefont
  {Ashcroft}}]{babaev2004superconductor}%
  \BibitemOpen
  \bibfield  {author} {\bibinfo {author} {\bibfnamefont {E.}~\bibnamefont
  {Babaev}}, \bibinfo {author} {\bibfnamefont {A.}~\bibnamefont {Sudb{\o}}},\
  and\ \bibinfo {author} {\bibfnamefont {N.}~\bibnamefont {Ashcroft}},\
  }\bibfield  {title} {\bibinfo {title} {A superconductor to superfluid phase
  transition in liquid metallic hydrogen},\ }\href
  {https://doi.org/10.1038/nature02910} {\bibfield  {journal} {\bibinfo
  {journal} {Nature}\ }\textbf {\bibinfo {volume} {431}},\ \bibinfo {pages}
  {666} (\bibinfo {year} {2004})}\BibitemShut {NoStop}%
\bibitem [{\citenamefont {Kuklov}\ \emph {et~al.}(2006)\citenamefont {Kuklov},
  \citenamefont {Prokof’Ev}, \citenamefont {Svistunov},\ and\ \citenamefont
  {Troyer}}]{kuklov2006deconfined}%
  \BibitemOpen
  \bibfield  {author} {\bibinfo {author} {\bibfnamefont {A.}~\bibnamefont
  {Kuklov}}, \bibinfo {author} {\bibfnamefont {N.}~\bibnamefont {Prokof’Ev}},
  \bibinfo {author} {\bibfnamefont {B.}~\bibnamefont {Svistunov}},\ and\
  \bibinfo {author} {\bibfnamefont {M.}~\bibnamefont {Troyer}},\ }\bibfield
  {title} {\bibinfo {title} {Deconfined criticality, runaway flow in the
  two-component scalar electrodynamics and weak first-order superfluid-solid
  transitions},\ }\href {https://doi.org/10.1016/j.aop.2006.04.007} {\bibfield
  {journal} {\bibinfo  {journal} {Annals of Physics}\ }\textbf {\bibinfo
  {volume} {321}},\ \bibinfo {pages} {1602} (\bibinfo {year}
  {2006})}\BibitemShut {NoStop}%
\bibitem [{\citenamefont {Babaev}(2004)}]{babaev2004phase}%
  \BibitemOpen
  \bibfield  {author} {\bibinfo {author} {\bibfnamefont {E.}~\bibnamefont
  {Babaev}},\ }\bibfield  {title} {\bibinfo {title} {Phase diagram of planar
  {U(1)$\times$ U(1)} superconductor: Condensation of vortices with fractional
  flux and a superfluid state},\ }\href
  {https://doi.org/10.1016/j.nuclphysb.2004.02.021} {\bibfield  {journal}
  {\bibinfo  {journal} {Nuclear Physics B}\ }\textbf {\bibinfo {volume}
  {686}},\ \bibinfo {pages} {397} (\bibinfo {year} {2004})}\BibitemShut
  {NoStop}%
\bibitem [{\citenamefont {Kuklov}\ and\ \citenamefont
  {Svistunov}(2003)}]{kuklov2003counterflow}%
  \BibitemOpen
  \bibfield  {author} {\bibinfo {author} {\bibfnamefont {A.~B.}\ \bibnamefont
  {Kuklov}}\ and\ \bibinfo {author} {\bibfnamefont {B.~V.}\ \bibnamefont
  {Svistunov}},\ }\bibfield  {title} {\bibinfo {title} {Counterflow
  superfluidity of two-species ultracold atoms in a commensurate optical
  lattice},\ }\href {https://doi.org/10.1103/PhysRevLett.90.100401} {\bibfield
  {journal} {\bibinfo  {journal} {Physical Review Letters}\ }\textbf {\bibinfo
  {volume} {90}},\ \bibinfo {pages} {100401} (\bibinfo {year}
  {2003})}\BibitemShut {NoStop}%
\bibitem [{\citenamefont {Kuklov}\ \emph {et~al.}(2004)\citenamefont {Kuklov},
  \citenamefont {Prokof'ev},\ and\ \citenamefont {Svistunov}}]{Kuklov2004}%
  \BibitemOpen
  \bibfield  {author} {\bibinfo {author} {\bibfnamefont {A.}~\bibnamefont
  {Kuklov}}, \bibinfo {author} {\bibfnamefont {N.}~\bibnamefont {Prokof'ev}},\
  and\ \bibinfo {author} {\bibfnamefont {B.}~\bibnamefont {Svistunov}},\
  }\bibfield  {title} {\bibinfo {title} {Superfluid-superfluid phase
  transitions in a two-component {Bose-Einstein} condensate},\ }\href
  {https://doi.org/10.1103/PhysRevLett.92.030403} {\bibfield  {journal}
  {\bibinfo  {journal} {Physical Review Letters}\ }\textbf {\bibinfo {volume}
  {92}},\ \bibinfo {pages} {030403} (\bibinfo {year} {2004})}\BibitemShut
  {NoStop}%
\bibitem [{\citenamefont {Herland}\ \emph {et~al.}(2010)\citenamefont
  {Herland}, \citenamefont {Babaev},\ and\ \citenamefont
  {Sudb\o{}}}]{Herland2010}%
  \BibitemOpen
  \bibfield  {author} {\bibinfo {author} {\bibfnamefont {E.~V.}\ \bibnamefont
  {Herland}}, \bibinfo {author} {\bibfnamefont {E.}~\bibnamefont {Babaev}},\
  and\ \bibinfo {author} {\bibfnamefont {A.}~\bibnamefont {Sudb\o{}}},\
  }\bibfield  {title} {\bibinfo {title} {Phase transitions in a three
  dimensional {U(1)$\times$ U(1)} lattice london superconductor: Metallic
  superfluid and charge-$4e$ superconducting states},\ }\href
  {https://doi.org/10.1103/PhysRevB.82.134511} {\bibfield  {journal} {\bibinfo
  {journal} {Physical Review B}\ }\textbf {\bibinfo {volume} {82}},\ \bibinfo
  {pages} {134511} (\bibinfo {year} {2010})}\BibitemShut {NoStop}%
\bibitem [{\citenamefont {Jian}\ \emph {et~al.}(2021)\citenamefont {Jian},
  \citenamefont {Huang},\ and\ \citenamefont {Yao}}]{Jian2021}%
  \BibitemOpen
  \bibfield  {author} {\bibinfo {author} {\bibfnamefont {S.-K.}\ \bibnamefont
  {Jian}}, \bibinfo {author} {\bibfnamefont {Y.}~\bibnamefont {Huang}},\ and\
  \bibinfo {author} {\bibfnamefont {H.}~\bibnamefont {Yao}},\ }\bibfield
  {title} {\bibinfo {title} {Charge-$4e$ superconductivity from nematic
  superconductors in two and three dimensions},\ }\href
  {https://doi.org/10.1103/PhysRevLett.127.227001} {\bibfield  {journal}
  {\bibinfo  {journal} {Physical Review Letters}\ }\textbf {\bibinfo {volume}
  {127}},\ \bibinfo {pages} {227001} (\bibinfo {year} {2021})}\BibitemShut
  {NoStop}%
\bibitem [{\citenamefont {Grinenko}\ \emph {et~al.}(2021)\citenamefont
  {Grinenko}, \citenamefont {Weston}, \citenamefont {Caglieris}, \citenamefont
  {Wuttke}, \citenamefont {Hess}, \citenamefont {Gottschall}, \citenamefont
  {Maccari}, \citenamefont {Gorbunov}, \citenamefont {Zherlitsyn},
  \citenamefont {Wosnitza}, \citenamefont {Rydh}, \citenamefont {Kihou},
  \citenamefont {Lee}, \citenamefont {Sarkar}, \citenamefont {Dengre},
  \citenamefont {Garaud}, \citenamefont {Charnukha}, \citenamefont {Hühne},
  \citenamefont {Nielsch}, \citenamefont {Büchner}, \citenamefont {Klauss},\
  and\ \citenamefont {Babaev}}]{Grinenko2021}%
  \BibitemOpen
  \bibfield  {author} {\bibinfo {author} {\bibfnamefont {V.}~\bibnamefont
  {Grinenko}}, \bibinfo {author} {\bibfnamefont {D.}~\bibnamefont {Weston}},
  \bibinfo {author} {\bibfnamefont {F.}~\bibnamefont {Caglieris}}, \bibinfo
  {author} {\bibfnamefont {C.}~\bibnamefont {Wuttke}}, \bibinfo {author}
  {\bibfnamefont {C.}~\bibnamefont {Hess}}, \bibinfo {author} {\bibfnamefont
  {T.}~\bibnamefont {Gottschall}}, \bibinfo {author} {\bibfnamefont
  {I.}~\bibnamefont {Maccari}}, \bibinfo {author} {\bibfnamefont
  {D.}~\bibnamefont {Gorbunov}}, \bibinfo {author} {\bibfnamefont
  {S.}~\bibnamefont {Zherlitsyn}}, \bibinfo {author} {\bibfnamefont
  {J.}~\bibnamefont {Wosnitza}}, \bibinfo {author} {\bibfnamefont
  {A.}~\bibnamefont {Rydh}}, \bibinfo {author} {\bibfnamefont {K.}~\bibnamefont
  {Kihou}}, \bibinfo {author} {\bibfnamefont {C.-H.}\ \bibnamefont {Lee}},
  \bibinfo {author} {\bibfnamefont {R.}~\bibnamefont {Sarkar}}, \bibinfo
  {author} {\bibfnamefont {S.}~\bibnamefont {Dengre}}, \bibinfo {author}
  {\bibfnamefont {J.}~\bibnamefont {Garaud}}, \bibinfo {author} {\bibfnamefont
  {A.}~\bibnamefont {Charnukha}}, \bibinfo {author} {\bibfnamefont
  {R.}~\bibnamefont {Hühne}}, \bibinfo {author} {\bibfnamefont
  {K.}~\bibnamefont {Nielsch}}, \bibinfo {author} {\bibfnamefont
  {B.}~\bibnamefont {Büchner}}, \bibinfo {author} {\bibfnamefont {H.-H.}\
  \bibnamefont {Klauss}},\ and\ \bibinfo {author} {\bibfnamefont
  {E.}~\bibnamefont {Babaev}},\ }\bibfield  {title} {\bibinfo {title} {State
  with spontaneously broken time-reversal symmetry above the superconducting
  phase transition},\ }\href {https://doi.org/10.1038/s41567-021-01350-9}
  {\bibfield  {journal} {\bibinfo  {journal} {Nature Physics}\ }\textbf
  {\bibinfo {volume} {17}},\ \bibinfo {pages} {1254} (\bibinfo {year}
  {2021})}\BibitemShut {NoStop}%
\bibitem [{\citenamefont {Cho}\ \emph {et~al.}(2022)\citenamefont {Cho},
  \citenamefont {Lyu}, \citenamefont {An}, \citenamefont {Han}, \citenamefont
  {Lo}, \citenamefont {Ng}, \citenamefont {Hu}, \citenamefont {Gao},
  \citenamefont {Li}, \citenamefont {Huang}, \citenamefont {Wang},
  \citenamefont {Schmalian},\ and\ \citenamefont {Lortz}}]{Cho2022}%
  \BibitemOpen
  \bibfield  {author} {\bibinfo {author} {\bibfnamefont {C.-w.}\ \bibnamefont
  {Cho}}, \bibinfo {author} {\bibfnamefont {J.}~\bibnamefont {Lyu}}, \bibinfo
  {author} {\bibfnamefont {L.}~\bibnamefont {An}}, \bibinfo {author}
  {\bibfnamefont {T.}~\bibnamefont {Han}}, \bibinfo {author} {\bibfnamefont
  {K.~T.}\ \bibnamefont {Lo}}, \bibinfo {author} {\bibfnamefont {C.~Y.}\
  \bibnamefont {Ng}}, \bibinfo {author} {\bibfnamefont {J.}~\bibnamefont {Hu}},
  \bibinfo {author} {\bibfnamefont {Y.}~\bibnamefont {Gao}}, \bibinfo {author}
  {\bibfnamefont {G.}~\bibnamefont {Li}}, \bibinfo {author} {\bibfnamefont
  {M.}~\bibnamefont {Huang}}, \bibinfo {author} {\bibfnamefont
  {N.}~\bibnamefont {Wang}}, \bibinfo {author} {\bibfnamefont {J.}~\bibnamefont
  {Schmalian}},\ and\ \bibinfo {author} {\bibfnamefont {R.}~\bibnamefont
  {Lortz}},\ }\bibfield  {title} {\bibinfo {title} {Nodal and nematic
  superconducting phases in {NbSe$_{2}$} monolayers from competing
  superconducting channels},\ }\href
  {https://doi.org/10.1103/PhysRevLett.129.087002} {\bibfield  {journal}
  {\bibinfo  {journal} {Physical Review Letters}\ }\textbf {\bibinfo {volume}
  {129}},\ \bibinfo {pages} {087002} (\bibinfo {year} {2022})}\BibitemShut
  {NoStop}%
\bibitem [{\citenamefont {Fernandes}\ \emph {et~al.}(2019)\citenamefont
  {Fernandes}, \citenamefont {Orth},\ and\ \citenamefont
  {Schmalian}}]{fernandes2019intertwined}%
  \BibitemOpen
  \bibfield  {author} {\bibinfo {author} {\bibfnamefont {R.~M.}\ \bibnamefont
  {Fernandes}}, \bibinfo {author} {\bibfnamefont {P.~P.}\ \bibnamefont
  {Orth}},\ and\ \bibinfo {author} {\bibfnamefont {J.}~\bibnamefont
  {Schmalian}},\ }\bibfield  {title} {\bibinfo {title} {Intertwined vestigial
  order in quantum materials: Nematicity and beyond},\ }\href
  {https://doi.org/10.1146/annurev-conmatphys-031218-013200} {\bibfield
  {journal} {\bibinfo  {journal} {Annual Review of Condensed Matter Physics}\
  }\textbf {\bibinfo {volume} {10}},\ \bibinfo {pages} {133} (\bibinfo {year}
  {2019})}\BibitemShut {NoStop}%
\bibitem [{\citenamefont {Aeppli}\ \emph {et~al.}(2020)\citenamefont {Aeppli},
  \citenamefont {Balatsky}, \citenamefont {R{\o}nnow},\ and\ \citenamefont
  {Spaldin}}]{aeppli2020hidden}%
  \BibitemOpen
  \bibfield  {author} {\bibinfo {author} {\bibfnamefont {G.}~\bibnamefont
  {Aeppli}}, \bibinfo {author} {\bibfnamefont {A.~V.}\ \bibnamefont
  {Balatsky}}, \bibinfo {author} {\bibfnamefont {H.~M.}\ \bibnamefont
  {R{\o}nnow}},\ and\ \bibinfo {author} {\bibfnamefont {N.~A.}\ \bibnamefont
  {Spaldin}},\ }\bibfield  {title} {\bibinfo {title} {Hidden, entangled and
  resonating order},\ }\href {https://doi.org/10.1038/s41578-020-0207-z}
  {\bibfield  {journal} {\bibinfo  {journal} {Nature Reviews Materials}\
  }\textbf {\bibinfo {volume} {5}},\ \bibinfo {pages} {477} (\bibinfo {year}
  {2020})}\BibitemShut {NoStop}%
\bibitem [{\citenamefont {Spaldin}\ \emph {et~al.}(2013)\citenamefont
  {Spaldin}, \citenamefont {Fechner}, \citenamefont {Bousquet}, \citenamefont
  {Balatsky},\ and\ \citenamefont {Nordstr\"om}}]{Spaldin2013}%
  \BibitemOpen
  \bibfield  {author} {\bibinfo {author} {\bibfnamefont {N.~A.}\ \bibnamefont
  {Spaldin}}, \bibinfo {author} {\bibfnamefont {M.}~\bibnamefont {Fechner}},
  \bibinfo {author} {\bibfnamefont {E.}~\bibnamefont {Bousquet}}, \bibinfo
  {author} {\bibfnamefont {A.}~\bibnamefont {Balatsky}},\ and\ \bibinfo
  {author} {\bibfnamefont {L.}~\bibnamefont {Nordstr\"om}},\ }\bibfield
  {title} {\bibinfo {title} {Monopole-based formalism for the diagonal
  magnetoelectric response},\ }\href
  {https://doi.org/10.1103/PhysRevB.88.094429} {\bibfield  {journal} {\bibinfo
  {journal} {Physical Review B}\ }\textbf {\bibinfo {volume} {88}},\ \bibinfo
  {pages} {094429} (\bibinfo {year} {2013})}\BibitemShut {NoStop}%
\bibitem [{\citenamefont {Th{\"o}le}\ and\ \citenamefont
  {Spaldin}(2018)}]{thole2018magnetoelectric}%
  \BibitemOpen
  \bibfield  {author} {\bibinfo {author} {\bibfnamefont {F.}~\bibnamefont
  {Th{\"o}le}}\ and\ \bibinfo {author} {\bibfnamefont {N.~A.}\ \bibnamefont
  {Spaldin}},\ }\bibfield  {title} {\bibinfo {title} {Magnetoelectric
  multipoles in metals},\ }\href {https://doi.org/10.1098/rsta.2017.0450}
  {\bibfield  {journal} {\bibinfo  {journal} {Philosophical Transactions of the
  Royal Society A: Mathematical, Physical and Engineering Sciences}\ }\textbf
  {\bibinfo {volume} {376}},\ \bibinfo {pages} {20170450} (\bibinfo {year}
  {2018})}\BibitemShut {NoStop}%
\bibitem [{\citenamefont {Geilhufe}\ and\ \citenamefont
  {Hergert}(2018)}]{gtpack1}%
  \BibitemOpen
  \bibfield  {author} {\bibinfo {author} {\bibfnamefont {R.~M.}\ \bibnamefont
  {Geilhufe}}\ and\ \bibinfo {author} {\bibfnamefont {W.}~\bibnamefont
  {Hergert}},\ }\bibfield  {title} {\bibinfo {title} {{GTPack: A Mathematica
  Group Theory Package for Application in Solid-State Physics and Photonics}},\
  }\href {https://doi.org/10.3389/fphy.2018.00086} {\bibfield  {journal}
  {\bibinfo  {journal} {Frontiers in Physics}\ }\textbf {\bibinfo {volume}
  {6}},\ \bibinfo {pages} {86} (\bibinfo {year} {2018})}\BibitemShut {NoStop}%
\bibitem [{\citenamefont {Barabanov}\ and\ \citenamefont
  {Kosogor}(2022)}]{barabanov2022landau}%
  \BibitemOpen
  \bibfield  {author} {\bibinfo {author} {\bibfnamefont {O.~V.}\ \bibnamefont
  {Barabanov}}\ and\ \bibinfo {author} {\bibfnamefont {A.}~\bibnamefont
  {Kosogor}},\ }\bibfield  {title} {\bibinfo {title} {Landau theory of
  ferroelastic phase transitions: Application to martensitic phase
  transformations},\ }\href
  {https://doi.org/https://doi.org/10.1063/10.0009538} {\bibfield  {journal}
  {\bibinfo  {journal} {Low Temperature Physics}\ }\textbf {\bibinfo {volume}
  {48}},\ \bibinfo {pages} {206} (\bibinfo {year} {2022})}\BibitemShut
  {NoStop}%
\bibitem [{\citenamefont {Devonshire}(1954)}]{devonshire1954theory}%
  \BibitemOpen
  \bibfield  {author} {\bibinfo {author} {\bibfnamefont {A.}~\bibnamefont
  {Devonshire}},\ }\bibfield  {title} {\bibinfo {title} {Theory of
  ferroelectrics},\ }\href {https://doi.org/10.1080/00018735400101173}
  {\bibfield  {journal} {\bibinfo  {journal} {Advances in physics}\ }\textbf
  {\bibinfo {volume} {3}},\ \bibinfo {pages} {85} (\bibinfo {year}
  {1954})}\BibitemShut {NoStop}%
\bibitem [{\citenamefont {Schwabl}(2006)}]{Schwabl2006}%
  \BibitemOpen
  \bibfield  {author} {\bibinfo {author} {\bibfnamefont {F.}~\bibnamefont
  {Schwabl}},\ }\href {https://doi.org/10.1007/3-540-36217-7} {\emph {\bibinfo
  {title} {Statistical Mechanics}}}\ (\bibinfo  {publisher} {Springer Berlin
  Heidelberg},\ \bibinfo {year} {2006})\BibitemShut {NoStop}%
\bibitem [{\citenamefont {Kleis}(1936)}]{Kleis1936}%
  \BibitemOpen
  \bibfield  {author} {\bibinfo {author} {\bibfnamefont {J.~D.}\ \bibnamefont
  {Kleis}},\ }\bibfield  {title} {\bibinfo {title} {Ferromagnetic anisotropy of
  nickel-iron crystals at various temperatures},\ }\href
  {https://doi.org/10.1103/PhysRev.50.1178} {\bibfield  {journal} {\bibinfo
  {journal} {Physical Review}\ }\textbf {\bibinfo {volume} {50}},\ \bibinfo
  {pages} {1178} (\bibinfo {year} {1936})}\BibitemShut {NoStop}%
\bibitem [{\citenamefont {van Vleck}(1937)}]{VanFleck1937}%
  \BibitemOpen
  \bibfield  {author} {\bibinfo {author} {\bibfnamefont {J.~H.}\ \bibnamefont
  {van Vleck}},\ }\bibfield  {title} {\bibinfo {title} {On the anisotropy of
  cubic ferromagnetic crystals},\ }\href
  {https://doi.org/10.1103/PhysRev.52.1178} {\bibfield  {journal} {\bibinfo
  {journal} {Physical Review}\ }\textbf {\bibinfo {volume} {52}},\ \bibinfo
  {pages} {1178} (\bibinfo {year} {1937})}\BibitemShut {NoStop}%
\bibitem [{\citenamefont {Altland}\ and\ \citenamefont
  {Simons}(2010{\natexlab{a}})}]{altland2010condensed}%
  \BibitemOpen
  \bibfield  {author} {\bibinfo {author} {\bibfnamefont {A.}~\bibnamefont
  {Altland}}\ and\ \bibinfo {author} {\bibfnamefont {B.~D.}\ \bibnamefont
  {Simons}},\ }\href@noop {} {\emph {\bibinfo {title} {Condensed matter field
  theory}}}\ (\bibinfo  {publisher} {Cambridge university press},\ \bibinfo
  {year} {2010})\BibitemShut {NoStop}%
\bibitem [{\citenamefont {Goto}\ \emph {et~al.}(2000)\citenamefont {Goto},
  \citenamefont {Nemoto}, \citenamefont {Nakano}, \citenamefont {Nakamura},
  \citenamefont {Kajitani},\ and\ \citenamefont {Kunii}}]{Goto2000}%
  \BibitemOpen
  \bibfield  {author} {\bibinfo {author} {\bibfnamefont {T.}~\bibnamefont
  {Goto}}, \bibinfo {author} {\bibfnamefont {Y.}~\bibnamefont {Nemoto}},
  \bibinfo {author} {\bibfnamefont {Y.}~\bibnamefont {Nakano}}, \bibinfo
  {author} {\bibfnamefont {S.}~\bibnamefont {Nakamura}}, \bibinfo {author}
  {\bibfnamefont {T.}~\bibnamefont {Kajitani}},\ and\ \bibinfo {author}
  {\bibfnamefont {S.}~\bibnamefont {Kunii}},\ }\bibfield  {title} {\bibinfo
  {title} {Quadrupolar effect of hob6 and dyb6},\ }\href
  {https://doi.org/10.1016/S0921-4526(99)01129-1} {\bibfield  {journal}
  {\bibinfo  {journal} {Physica B: Condensed Matter}\ }\textbf {\bibinfo
  {volume} {281-282}},\ \bibinfo {pages} {586} (\bibinfo {year}
  {2000})}\BibitemShut {NoStop}%
\bibitem [{\citenamefont {Marjerrison}\ \emph {et~al.}(2016)\citenamefont
  {Marjerrison}, \citenamefont {Thompson}, \citenamefont {Sala}, \citenamefont
  {Maharaj}, \citenamefont {Kermarrec}, \citenamefont {Cai}, \citenamefont
  {Hallas}, \citenamefont {Wilson}, \citenamefont {Munsie}, \citenamefont
  {Granroth}, \citenamefont {Flacau}, \citenamefont {Greedan}, \citenamefont
  {Gaulin},\ and\ \citenamefont {Luke}}]{marjerrison2016cubic}%
  \BibitemOpen
  \bibfield  {author} {\bibinfo {author} {\bibfnamefont {C.~A.}\ \bibnamefont
  {Marjerrison}}, \bibinfo {author} {\bibfnamefont {C.~M.}\ \bibnamefont
  {Thompson}}, \bibinfo {author} {\bibfnamefont {G.}~\bibnamefont {Sala}},
  \bibinfo {author} {\bibfnamefont {D.~D.}\ \bibnamefont {Maharaj}}, \bibinfo
  {author} {\bibfnamefont {E.}~\bibnamefont {Kermarrec}}, \bibinfo {author}
  {\bibfnamefont {Y.}~\bibnamefont {Cai}}, \bibinfo {author} {\bibfnamefont
  {A.~M.}\ \bibnamefont {Hallas}}, \bibinfo {author} {\bibfnamefont {M.~N.}\
  \bibnamefont {Wilson}}, \bibinfo {author} {\bibfnamefont {T.~J.~S.}\
  \bibnamefont {Munsie}}, \bibinfo {author} {\bibfnamefont {G.~E.}\
  \bibnamefont {Granroth}}, \bibinfo {author} {\bibfnamefont {R.}~\bibnamefont
  {Flacau}}, \bibinfo {author} {\bibfnamefont {J.~E.}\ \bibnamefont {Greedan}},
  \bibinfo {author} {\bibfnamefont {B.~D.}\ \bibnamefont {Gaulin}},\ and\
  \bibinfo {author} {\bibfnamefont {G.~M.}\ \bibnamefont {Luke}},\ }\bibfield
  {title} {\bibinfo {title} {{Cubic Re$^{6+}$ (5d$^1$) Double Perovskites,
  {Ba$_2$MgReO$_6$, Ba$_2$ZnReO$_6$, and Ba$_2$Y$_{\frac{2}{3}}$ReO$_6$}:
  Magnetism, Heat Capacity, {$\mu$SR}, and Neutron Scattering Studies and
  Comparison with Theory}},\ }\href
  {https://doi.org/10.1021/acs.inorgchem.6b01933} {\bibfield  {journal}
  {\bibinfo  {journal} {Inorganic Chemistry}\ }\textbf {\bibinfo {volume}
  {55}},\ \bibinfo {pages} {10701} (\bibinfo {year} {2016})}\BibitemShut
  {NoStop}%
\bibitem [{\citenamefont {Hirai}\ \emph {et~al.}(2020)\citenamefont {Hirai},
  \citenamefont {Sagayama}, \citenamefont {Gao}, \citenamefont {Ohsumi},
  \citenamefont {Chen}, \citenamefont {Arima},\ and\ \citenamefont
  {Hiroi}}]{Hirai2020}%
  \BibitemOpen
  \bibfield  {author} {\bibinfo {author} {\bibfnamefont {D.}~\bibnamefont
  {Hirai}}, \bibinfo {author} {\bibfnamefont {H.}~\bibnamefont {Sagayama}},
  \bibinfo {author} {\bibfnamefont {S.}~\bibnamefont {Gao}}, \bibinfo {author}
  {\bibfnamefont {H.}~\bibnamefont {Ohsumi}}, \bibinfo {author} {\bibfnamefont
  {G.}~\bibnamefont {Chen}}, \bibinfo {author} {\bibfnamefont {T.-h.}\
  \bibnamefont {Arima}},\ and\ \bibinfo {author} {\bibfnamefont
  {Z.}~\bibnamefont {Hiroi}},\ }\bibfield  {title} {\bibinfo {title} {Detection
  of multipolar orders in the spin-orbit-coupled $5d$ mott insulator
  $\mathrm{B}{\mathrm{a}}_{2}\mathrm{MgRe}{\mathrm{o}}_{6}$},\ }\href
  {https://doi.org/10.1103/PhysRevResearch.2.022063} {\bibfield  {journal}
  {\bibinfo  {journal} {Physical Review Research}\ }\textbf {\bibinfo {volume}
  {2}},\ \bibinfo {pages} {022063} (\bibinfo {year} {2020})}\BibitemShut
  {NoStop}%
\bibitem [{\citenamefont {P{\'a}sztorov{\'a}}\ \emph
  {et~al.}(2023)\citenamefont {P{\'a}sztorov{\'a}}, \citenamefont
  {Mansouri~Tehrani}, \citenamefont {{\v{Z}}ivkovi{\'c}}, \citenamefont
  {Spaldin},\ and\ \citenamefont {R{\o}nnow}}]{pasztorova2023experimental}%
  \BibitemOpen
  \bibfield  {author} {\bibinfo {author} {\bibfnamefont {J.}~\bibnamefont
  {P{\'a}sztorov{\'a}}}, \bibinfo {author} {\bibfnamefont {A.}~\bibnamefont
  {Mansouri~Tehrani}}, \bibinfo {author} {\bibfnamefont {I.}~\bibnamefont
  {{\v{Z}}ivkovi{\'c}}}, \bibinfo {author} {\bibfnamefont {N.~A.}\ \bibnamefont
  {Spaldin}},\ and\ \bibinfo {author} {\bibfnamefont {H.~M.}\ \bibnamefont
  {R{\o}nnow}},\ }\bibfield  {title} {\bibinfo {title} {Experimental and
  theoretical thermodynamic studies in {Ba$_2$MgReO$_6$}—the ground state in
  the context of {Jahn-Teller effect}},\ }\href
  {https://doi.org/10.1088/1361-648X/acc62a} {\bibfield  {journal} {\bibinfo
  {journal} {Journal of Physics: Condensed Matter}\ }\textbf {\bibinfo {volume}
  {35}},\ \bibinfo {pages} {245603} (\bibinfo {year} {2023})}\BibitemShut
  {NoStop}%
\bibitem [{\citenamefont {Ketfi}\ \emph {et~al.}(2024)\citenamefont {Ketfi},
  \citenamefont {Essaoud}, \citenamefont {Al~Azar}, \citenamefont {Al-Reyahi},
  \citenamefont {Mousa},\ and\ \citenamefont
  {Al-Aqtash}}]{ketfi2024mechanical}%
  \BibitemOpen
  \bibfield  {author} {\bibinfo {author} {\bibfnamefont {M.~E.}\ \bibnamefont
  {Ketfi}}, \bibinfo {author} {\bibfnamefont {S.~S.}\ \bibnamefont {Essaoud}},
  \bibinfo {author} {\bibfnamefont {S.~M.}\ \bibnamefont {Al~Azar}}, \bibinfo
  {author} {\bibfnamefont {A.~Y.}\ \bibnamefont {Al-Reyahi}}, \bibinfo {author}
  {\bibfnamefont {A.~A.}\ \bibnamefont {Mousa}},\ and\ \bibinfo {author}
  {\bibfnamefont {N.}~\bibnamefont {Al-Aqtash}},\ }\bibfield  {title} {\bibinfo
  {title} {Mechanical, magneto-electronic and thermoelectric properties of
  ba$_2$mgreo$_6$ and ba$_2$ymoo$_6$ based cubic double perovskites: an ab
  initio study},\ }\href {https://doi.org/10.1088/1402-4896/ad1021} {\bibfield
  {journal} {\bibinfo  {journal} {Physica Scripta}\ }\textbf {\bibinfo {volume}
  {99}},\ \bibinfo {pages} {015908} (\bibinfo {year} {2024})}\BibitemShut
  {NoStop}%
\bibitem [{\citenamefont {Mansouri~Tehrani}\ and\ \citenamefont
  {Spaldin}(2021)}]{Tehrani2021}%
  \BibitemOpen
  \bibfield  {author} {\bibinfo {author} {\bibfnamefont {A.}~\bibnamefont
  {Mansouri~Tehrani}}\ and\ \bibinfo {author} {\bibfnamefont {N.~A.}\
  \bibnamefont {Spaldin}},\ }\bibfield  {title} {\bibinfo {title} {Untangling
  the structural, magnetic dipole, and charge multipolar orders in
  ${\mathrm{ba}}_{2}{\mathrm{mgreo}}_{6}$},\ }\href
  {https://doi.org/10.1103/PhysRevMaterials.5.104410} {\bibfield  {journal}
  {\bibinfo  {journal} {Physical Review Materials}\ }\textbf {\bibinfo {volume}
  {5}},\ \bibinfo {pages} {104410} (\bibinfo {year} {2021})}\BibitemShut
  {NoStop}%
\bibitem [{\citenamefont {Ishikawa}\ \emph {et~al.}(2020)\citenamefont
  {Ishikawa}, \citenamefont {Yajima}, \citenamefont {Matsuo}, \citenamefont
  {Ihara},\ and\ \citenamefont {Kindo}}]{Ishikawa2020}%
  \BibitemOpen
  \bibfield  {author} {\bibinfo {author} {\bibfnamefont {H.}~\bibnamefont
  {Ishikawa}}, \bibinfo {author} {\bibfnamefont {T.}~\bibnamefont {Yajima}},
  \bibinfo {author} {\bibfnamefont {A.}~\bibnamefont {Matsuo}}, \bibinfo
  {author} {\bibfnamefont {Y.}~\bibnamefont {Ihara}},\ and\ \bibinfo {author}
  {\bibfnamefont {K.}~\bibnamefont {Kindo}},\ }\bibfield  {title} {\bibinfo
  {title} {Nonmagnetic ground states and a possible quadrupolar phase in $4d$
  and $5d$ lacunar spinel selenides {GaM$_4$Se$_8$} ({M= Nb, Ta})},\ }\href
  {https://doi.org/10.1103/PhysRevLett.124.227202} {\bibfield  {journal}
  {\bibinfo  {journal} {Physical Review Letters}\ }\textbf {\bibinfo {volume}
  {124}},\ \bibinfo {pages} {227202} (\bibinfo {year} {2020})}\BibitemShut
  {NoStop}%
\bibitem [{\citenamefont {Trinh}\ \emph {et~al.}(2016)\citenamefont {Trinh},
  \citenamefont {Br\"uck}, \citenamefont {Siegrist}, \citenamefont {Flint},
  \citenamefont {Chandra}, \citenamefont {Coleman},\ and\ \citenamefont
  {Ramirez}}]{Trinh2016}%
  \BibitemOpen
  \bibfield  {author} {\bibinfo {author} {\bibfnamefont {J.}~\bibnamefont
  {Trinh}}, \bibinfo {author} {\bibfnamefont {E.}~\bibnamefont {Br\"uck}},
  \bibinfo {author} {\bibfnamefont {T.}~\bibnamefont {Siegrist}}, \bibinfo
  {author} {\bibfnamefont {R.}~\bibnamefont {Flint}}, \bibinfo {author}
  {\bibfnamefont {P.}~\bibnamefont {Chandra}}, \bibinfo {author} {\bibfnamefont
  {P.}~\bibnamefont {Coleman}},\ and\ \bibinfo {author} {\bibfnamefont {A.~P.}\
  \bibnamefont {Ramirez}},\ }\bibfield  {title} {\bibinfo {title}
  {Thermodynamic measurement of angular anisotropy at the hidden order
  transition of {URu$_2$Si$_2$}},\ }\href
  {https://doi.org/10.1103/PhysRevLett.117.157201} {\bibfield  {journal}
  {\bibinfo  {journal} {Physical Review Letters}\ }\textbf {\bibinfo {volume}
  {117}},\ \bibinfo {pages} {157201} (\bibinfo {year} {2016})}\BibitemShut
  {NoStop}%
\bibitem [{\citenamefont {Brassington}\ and\ \citenamefont
  {Saunders}(1982)}]{Brassington1982}%
  \BibitemOpen
  \bibfield  {author} {\bibinfo {author} {\bibfnamefont {M.~P.}\ \bibnamefont
  {Brassington}}\ and\ \bibinfo {author} {\bibfnamefont {G.~A.}\ \bibnamefont
  {Saunders}},\ }\bibfield  {title} {\bibinfo {title} {Cubic invariants in the
  {Landau} theory applied to elastic phase transitions},\ }\href
  {https://doi.org/10.1103/PhysRevLett.48.159} {\bibfield  {journal} {\bibinfo
  {journal} {Physical Review Letters}\ }\textbf {\bibinfo {volume} {48}},\
  \bibinfo {pages} {159} (\bibinfo {year} {1982})}\BibitemShut {NoStop}%
\bibitem [{\citenamefont {Aktas}\ \emph {et~al.}(2013)\citenamefont {Aktas},
  \citenamefont {Salje}, \citenamefont {Crossley}, \citenamefont {Lampronti},
  \citenamefont {Whatmore}, \citenamefont {Mathur},\ and\ \citenamefont
  {Carpenter}}]{Aktas2013}%
  \BibitemOpen
  \bibfield  {author} {\bibinfo {author} {\bibfnamefont {O.}~\bibnamefont
  {Aktas}}, \bibinfo {author} {\bibfnamefont {E.~K.~H.}\ \bibnamefont {Salje}},
  \bibinfo {author} {\bibfnamefont {S.}~\bibnamefont {Crossley}}, \bibinfo
  {author} {\bibfnamefont {G.~I.}\ \bibnamefont {Lampronti}}, \bibinfo {author}
  {\bibfnamefont {R.~W.}\ \bibnamefont {Whatmore}}, \bibinfo {author}
  {\bibfnamefont {N.~D.}\ \bibnamefont {Mathur}},\ and\ \bibinfo {author}
  {\bibfnamefont {M.~A.}\ \bibnamefont {Carpenter}},\ }\bibfield  {title}
  {\bibinfo {title} {Ferroelectric precursor behavior in
  {PbSc$_{0.5}$Ta$_{0.5}$O$_{3}$} detected by field-induced resonant
  piezoelectric spectroscopy},\ }\href
  {https://doi.org/10.1103/PhysRevB.88.174112} {\bibfield  {journal} {\bibinfo
  {journal} {Physical Review B}\ }\textbf {\bibinfo {volume} {88}},\ \bibinfo
  {pages} {174112} (\bibinfo {year} {2013})}\BibitemShut {NoStop}%
\bibitem [{\citenamefont {Fiebig}(2005)}]{fiebig2005revival}%
  \BibitemOpen
  \bibfield  {author} {\bibinfo {author} {\bibfnamefont {M.}~\bibnamefont
  {Fiebig}},\ }\bibfield  {title} {\bibinfo {title} {Revival of the
  magnetoelectric effect},\ }\href {https://doi.org/10.1088/0022-3727/38/8/R01}
  {\bibfield  {journal} {\bibinfo  {journal} {Journal of physics D: applied
  physics}\ }\textbf {\bibinfo {volume} {38}},\ \bibinfo {pages} {R123}
  (\bibinfo {year} {2005})}\BibitemShut {NoStop}%
\bibitem [{\citenamefont {Khomskii}(2006)}]{KHOMSKII20061}%
  \BibitemOpen
  \bibfield  {author} {\bibinfo {author} {\bibfnamefont {D.}~\bibnamefont
  {Khomskii}},\ }\bibfield  {title} {\bibinfo {title} {Multiferroics: Different
  ways to combine magnetism and ferroelectricity},\ }\href
  {https://doi.org/https://doi.org/10.1016/j.jmmm.2006.01.238} {\bibfield
  {journal} {\bibinfo  {journal} {Journal of Magnetism and Magnetic Materials}\
  }\textbf {\bibinfo {volume} {306}},\ \bibinfo {pages} {1} (\bibinfo {year}
  {2006})}\BibitemShut {NoStop}%
\bibitem [{\citenamefont {Ramesh}\ and\ \citenamefont
  {Spaldin}(2007)}]{Ramesh2007}%
  \BibitemOpen
  \bibfield  {author} {\bibinfo {author} {\bibfnamefont {R.}~\bibnamefont
  {Ramesh}}\ and\ \bibinfo {author} {\bibfnamefont {N.~A.}\ \bibnamefont
  {Spaldin}},\ }\bibfield  {title} {\bibinfo {title} {Multiferroics: progress
  and prospects in thin films},\ }\href {https://doi.org/10.1038/nmat1805}
  {\bibfield  {journal} {\bibinfo  {journal} {Nature Materials}\ }\textbf
  {\bibinfo {volume} {6}},\ \bibinfo {pages} {21} (\bibinfo {year}
  {2007})}\BibitemShut {NoStop}%
\bibitem [{\citenamefont {Fiebig}\ \emph {et~al.}(2016)\citenamefont {Fiebig},
  \citenamefont {Lottermoser}, \citenamefont {Meier},\ and\ \citenamefont
  {Trassin}}]{Fiebig2016}%
  \BibitemOpen
  \bibfield  {author} {\bibinfo {author} {\bibfnamefont {M.}~\bibnamefont
  {Fiebig}}, \bibinfo {author} {\bibfnamefont {T.}~\bibnamefont {Lottermoser}},
  \bibinfo {author} {\bibfnamefont {D.}~\bibnamefont {Meier}},\ and\ \bibinfo
  {author} {\bibfnamefont {M.}~\bibnamefont {Trassin}},\ }\bibfield  {title}
  {\bibinfo {title} {The evolution of multiferroics},\ }\href
  {https://doi.org/10.1038/natrevmats.2016.46} {\bibfield  {journal} {\bibinfo
  {journal} {Nature Reviews Materials}\ }\textbf {\bibinfo {volume} {1}},\
  \bibinfo {pages} {16046} (\bibinfo {year} {2016})}\BibitemShut {NoStop}%
\bibitem [{\citenamefont {Tokura}\ \emph {et~al.}(2014)\citenamefont {Tokura},
  \citenamefont {Seki},\ and\ \citenamefont {Nagaosa}}]{Tokura_2014}%
  \BibitemOpen
  \bibfield  {author} {\bibinfo {author} {\bibfnamefont {Y.}~\bibnamefont
  {Tokura}}, \bibinfo {author} {\bibfnamefont {S.}~\bibnamefont {Seki}},\ and\
  \bibinfo {author} {\bibfnamefont {N.}~\bibnamefont {Nagaosa}},\ }\bibfield
  {title} {\bibinfo {title} {Multiferroics of spin origin},\ }\href
  {https://doi.org/10.1088/0034-4885/77/7/076501} {\bibfield  {journal}
  {\bibinfo  {journal} {Reports on Progress in Physics}\ }\textbf {\bibinfo
  {volume} {77}},\ \bibinfo {pages} {076501} (\bibinfo {year}
  {2014})}\BibitemShut {NoStop}%
\bibitem [{\citenamefont {Spaldin}\ and\ \citenamefont
  {Ramesh}(2019)}]{spaldin2019}%
  \BibitemOpen
  \bibfield  {author} {\bibinfo {author} {\bibfnamefont {N.~A.}\ \bibnamefont
  {Spaldin}}\ and\ \bibinfo {author} {\bibfnamefont {R.}~\bibnamefont
  {Ramesh}},\ }\bibfield  {title} {\bibinfo {title} {Advances in
  magnetoelectric multiferroics},\ }\href
  {https://doi.org/10.1038/s41563-018-0275-2} {\bibfield  {journal} {\bibinfo
  {journal} {Nature Materials}\ }\textbf {\bibinfo {volume} {18}},\ \bibinfo
  {pages} {203} (\bibinfo {year} {2019})}\BibitemShut {NoStop}%
\bibitem [{\citenamefont {Bossini}\ \emph {et~al.}(2023)\citenamefont
  {Bossini}, \citenamefont {Juraschek}, \citenamefont {Geilhufe}, \citenamefont
  {Nagaosa}, \citenamefont {Balatsky}, \citenamefont {Milanovi{\'c}},
  \citenamefont {Srdi{\'c}}, \citenamefont {{\v{S}}enjug}, \citenamefont
  {Topi{\'c}}, \citenamefont {Bari{\v{s}}i{\'c}} \emph
  {et~al.}}]{bossini2023magnetoelectrics}%
  \BibitemOpen
  \bibfield  {author} {\bibinfo {author} {\bibfnamefont {D.}~\bibnamefont
  {Bossini}}, \bibinfo {author} {\bibfnamefont {D.~M.}\ \bibnamefont
  {Juraschek}}, \bibinfo {author} {\bibfnamefont {R.~M.}\ \bibnamefont
  {Geilhufe}}, \bibinfo {author} {\bibfnamefont {N.}~\bibnamefont {Nagaosa}},
  \bibinfo {author} {\bibfnamefont {A.~V.}\ \bibnamefont {Balatsky}}, \bibinfo
  {author} {\bibfnamefont {M.}~\bibnamefont {Milanovi{\'c}}}, \bibinfo {author}
  {\bibfnamefont {V.~V.}\ \bibnamefont {Srdi{\'c}}}, \bibinfo {author}
  {\bibfnamefont {P.}~\bibnamefont {{\v{S}}enjug}}, \bibinfo {author}
  {\bibfnamefont {E.}~\bibnamefont {Topi{\'c}}}, \bibinfo {author}
  {\bibfnamefont {D.}~\bibnamefont {Bari{\v{s}}i{\'c}}}, \emph {et~al.},\
  }\bibfield  {title} {\bibinfo {title} {Magnetoelectrics and multiferroics:
  theory, synthesis, characterisation, preliminary results and perspectives for
  all-optical manipulations},\ }\href
  {https://doi.org/10.1088/1361-6463/acc8e1} {\bibfield  {journal} {\bibinfo
  {journal} {Journal of Physics D: Applied Physics}\ }\textbf {\bibinfo
  {volume} {56}},\ \bibinfo {pages} {273001} (\bibinfo {year}
  {2023})}\BibitemShut {NoStop}%
\bibitem [{\citenamefont {Wang}\ \emph {et~al.}(2003)\citenamefont {Wang},
  \citenamefont {Neaton}, \citenamefont {Zheng}, \citenamefont {Nagarajan},
  \citenamefont {Ogale}, \citenamefont {Liu}, \citenamefont {Viehland},
  \citenamefont {Vaithyanathan}, \citenamefont {Schlom}, \citenamefont
  {Waghmare}, \citenamefont {Spaldin}, \citenamefont {Rabe}, \citenamefont
  {Wuttig},\ and\ \citenamefont {Ramesh}}]{Wang2003}%
  \BibitemOpen
  \bibfield  {author} {\bibinfo {author} {\bibfnamefont {J.}~\bibnamefont
  {Wang}}, \bibinfo {author} {\bibfnamefont {J.~B.}\ \bibnamefont {Neaton}},
  \bibinfo {author} {\bibfnamefont {H.}~\bibnamefont {Zheng}}, \bibinfo
  {author} {\bibfnamefont {V.}~\bibnamefont {Nagarajan}}, \bibinfo {author}
  {\bibfnamefont {S.~B.}\ \bibnamefont {Ogale}}, \bibinfo {author}
  {\bibfnamefont {B.}~\bibnamefont {Liu}}, \bibinfo {author} {\bibfnamefont
  {D.}~\bibnamefont {Viehland}}, \bibinfo {author} {\bibfnamefont
  {V.}~\bibnamefont {Vaithyanathan}}, \bibinfo {author} {\bibfnamefont {D.~G.}\
  \bibnamefont {Schlom}}, \bibinfo {author} {\bibfnamefont {U.~V.}\
  \bibnamefont {Waghmare}}, \bibinfo {author} {\bibfnamefont {N.~A.}\
  \bibnamefont {Spaldin}}, \bibinfo {author} {\bibfnamefont {K.~M.}\
  \bibnamefont {Rabe}}, \bibinfo {author} {\bibfnamefont {M.}~\bibnamefont
  {Wuttig}},\ and\ \bibinfo {author} {\bibfnamefont {R.}~\bibnamefont
  {Ramesh}},\ }\bibfield  {title} {\bibinfo {title} {Epitaxial {BiFeO$_3$}
  multiferroic thin film heterostructures},\ }\href
  {https://doi.org/10.1126/science.1080615} {\bibfield  {journal} {\bibinfo
  {journal} {Science}\ }\textbf {\bibinfo {volume} {299}},\ \bibinfo {pages}
  {1719} (\bibinfo {year} {2003})}\BibitemShut {NoStop}%
\bibitem [{\citenamefont {Lee}\ \emph {et~al.}(2015)\citenamefont {Lee},
  \citenamefont {Kim}, \citenamefont {Park}, \citenamefont {Kim}, \citenamefont
  {Song}, \citenamefont {Kim}, \citenamefont {Kim}, \citenamefont {Do},\ and\
  \citenamefont {Jeong}}]{Lee2015}%
  \BibitemOpen
  \bibfield  {author} {\bibinfo {author} {\bibfnamefont {M.~H.}\ \bibnamefont
  {Lee}}, \bibinfo {author} {\bibfnamefont {D.~J.}\ \bibnamefont {Kim}},
  \bibinfo {author} {\bibfnamefont {J.~S.}\ \bibnamefont {Park}}, \bibinfo
  {author} {\bibfnamefont {S.~W.}\ \bibnamefont {Kim}}, \bibinfo {author}
  {\bibfnamefont {T.~K.}\ \bibnamefont {Song}}, \bibinfo {author}
  {\bibfnamefont {M.-H.}\ \bibnamefont {Kim}}, \bibinfo {author} {\bibfnamefont
  {W.-J.}\ \bibnamefont {Kim}}, \bibinfo {author} {\bibfnamefont
  {D.}~\bibnamefont {Do}},\ and\ \bibinfo {author} {\bibfnamefont {I.-K.}\
  \bibnamefont {Jeong}},\ }\bibfield  {title} {\bibinfo {title}
  {High-performance lead-free piezoceramics with high curie temperatures},\
  }\href {https://doi.org/https://doi.org/10.1002/adma.201502424} {\bibfield
  {journal} {\bibinfo  {journal} {Advanced Materials}\ }\textbf {\bibinfo
  {volume} {27}},\ \bibinfo {pages} {6976} (\bibinfo {year}
  {2015})}\BibitemShut {NoStop}%
\bibitem [{\citenamefont {Kung}\ \emph {et~al.}(2016)\citenamefont {Kung},
  \citenamefont {Ran}, \citenamefont {Kanchanavatee}, \citenamefont {Krapivin},
  \citenamefont {Lee}, \citenamefont {Mydosh}, \citenamefont {Haule},
  \citenamefont {Maple},\ and\ \citenamefont {Blumberg}}]{Kung2016}%
  \BibitemOpen
  \bibfield  {author} {\bibinfo {author} {\bibfnamefont {H.-H.}\ \bibnamefont
  {Kung}}, \bibinfo {author} {\bibfnamefont {S.}~\bibnamefont {Ran}}, \bibinfo
  {author} {\bibfnamefont {N.}~\bibnamefont {Kanchanavatee}}, \bibinfo {author}
  {\bibfnamefont {V.}~\bibnamefont {Krapivin}}, \bibinfo {author}
  {\bibfnamefont {A.}~\bibnamefont {Lee}}, \bibinfo {author} {\bibfnamefont
  {J.~A.}\ \bibnamefont {Mydosh}}, \bibinfo {author} {\bibfnamefont
  {K.}~\bibnamefont {Haule}}, \bibinfo {author} {\bibfnamefont {M.~B.}\
  \bibnamefont {Maple}},\ and\ \bibinfo {author} {\bibfnamefont
  {G.}~\bibnamefont {Blumberg}},\ }\bibfield  {title} {\bibinfo {title}
  {Analogy between the ``hidden order'' and the orbital antiferromagnetism in
  {URu$_{2-x}$Fe$_x$Si$_{2}$}},\ }\href
  {https://doi.org/10.1103/PhysRevLett.117.227601} {\bibfield  {journal}
  {\bibinfo  {journal} {Physical Review Letters}\ }\textbf {\bibinfo {volume}
  {117}},\ \bibinfo {pages} {227601} (\bibinfo {year} {2016})}\BibitemShut
  {NoStop}%
\bibitem [{\citenamefont {Altland}\ and\ \citenamefont
  {Simons}(2010{\natexlab{b}})}]{Altland2010}%
  \BibitemOpen
  \bibfield  {author} {\bibinfo {author} {\bibfnamefont {A.}~\bibnamefont
  {Altland}}\ and\ \bibinfo {author} {\bibfnamefont {B.~D.}\ \bibnamefont
  {Simons}},\ }\href {https://doi.org/10.1017/CBO9780511789984} {\emph
  {\bibinfo {title} {Condensed Matter Field Theory}}}\ (\bibinfo  {publisher}
  {Cambridge University Press},\ \bibinfo {year} {2010})\BibitemShut {NoStop}%
\end{thebibliography}
\end{document}